%% file: _main.tex
\documentclass[sigconf, natbib=false]{acmart}
\AtBeginDocument{%
  }

\usepackage{todonotes}

\setcopyright{acmlicensed}
\copyrightyear{2026}
\acmYear{2026}
\acmDOI{XXXXXXX.XXXXXXX}
\acmConference[MM '26]{the 34th ACM International Conference on Multimedia}{November 10--14, 2026}{Rio de Janeiro, Brazil}
\acmISBN{978-1-4503-XXXX-X/2018/06}

\acmSubmissionID{9581}

\RequirePackage[
  datamodel=acmdatamodel,
  style=acmnumeric,
  ]{biblatex}

\input{_macros}

\begin{document}
\title{GenGA: Editable and Data-Grounded Graphical Abstract Generation for Academic Papers}

\author{Takuro Kawada}
\orcid{0009-0004-3142-6108}
\affiliation{%
  \institution{Hosei University}
  \city{Tokyo}
  \country{Japan}
}
\email{takuro.kawada.3g@stu.hosei.ac.jp}

\author{Shunsuke Kitada}
\orcid{0000-0002-3330-8779}
\affiliation{%
  \institution{Hosei University}
  \city{Tokyo}
  \country{Japan}
}
\email{info@shunk031.me}

\author{Hitoshi Iyatomi}
\orcid{0000-0003-4108-4178}
\affiliation{%
  \institution{Hosei University}
  \city{Tokyo}
  \country{Japan}
}
\email{iyatomi@hosei.ac.jp}

\renewcommand{\shortauthors}{Kawada et al.}

\input{sec/00_abstract}

\ccsdesc[500]{Information systems~Multimedia content creation}
\ccsdesc[500]{Applied computing~Multi / mixed media creation}
\ccsdesc[500]{Applied computing~Image composition}
\ccsdesc[300]{Computing methodologies~Computer vision}
\ccsdesc[300]{Computing methodologies~Natural language processing}

\keywords{Graphical Abstracts, Visual Summarization, Editable Generation, Vector Graphics, Vision-Language Models}
\input{fig/tex/01_teaser}


\maketitle

\input{sec/01_introduction}
\input{sec/02_related_work}
\input{sec/03_proposed}
\input{sec/04_experiments}
\input{sec/05_results_and_discussion}
\input{sec/06_conclusion}

\clearpage
\printbibliography

\clearpage
\appendix
\twocolumn[
\begin{center}
    {\LARGE \textbf{GenGA: Editable and Data-Grounded Graphical Abstract Generation}\par}
    \vspace{1em}
    {\large Supplementary Material\par}
    \vspace{2em}
\end{center}
]

\input{sec/appendix/01_dataset_curation}
\input{sec/appendix/02_prompt}
\input{sec/appendix/03_additional_results}
\input{sec/appendix/04_user_study_setup}

\input{sec/appendix/_materials}

\end{document}

%% file: _macros.tex
\usepackage{graphicx}	
\usepackage{amsmath}
\usepackage{booktabs}
\usepackage{times}
\usepackage{microtype}
\usepackage{epsfig}
\usepackage{caption}
\usepackage{float}
\usepackage{placeins}
\usepackage{color, colortbl}
\usepackage{stfloats}
\usepackage{enumitem}
\usepackage{tabularx}
\usepackage{xstring}
\usepackage{multirow}
\usepackage{xspace}
\usepackage{url}
\usepackage{subcaption}
\usepackage{color}
\usepackage{xcolor}
\usepackage[hang,flushmargin]{footmisc}
\usepackage{pifont}
\usepackage{listings}
\usepackage[most]{tcolorbox}
\usepackage{fvextra}
\usepackage{pifont}
\usepackage{hyperref}
\usepackage{mathtools}

\newcommand{\cmark}{\textcolor{green!60!black}{\ding{51}}}
\newcommand{\xmark}{\textcolor{red!80!black}{\ding{55}}} 

\definecolor{promptbg}{RGB}{248,248,248}
\definecolor{promptframe}{RGB}{210,210,210}

\DefineVerbatimEnvironment{PromptVerb}{Verbatim}{
  breaklines=true,
  breakanywhere=true,
  breaksymbolleft={},
  breaksymbolright={},
  fontsize=\footnotesize,
  numbers=left,
  numbersep=6pt
}

\newcounter{promptboxctr}

\newtcolorbox[auto counter]{promptbox}[3][blue]{%
    enhanced,
    breakable,
    colback=white,
    colframe=#1,
    colbacktitle=#1!10,
    coltitle=black,
    fonttitle=\bfseries\sffamily,
    boxrule=0.8pt,
    title={Prompt~\thetcbcounter: #2},
    label={#3}
}

%% file: sec/00_abstract.tex
\begin{abstract}

Graphical Abstracts (GAs) visually summarize the key findings of academic papers, playing a crucial role in facilitating the understanding of research content.
Recently, advancements in vision-language models and image generation models have enabled the automatic generation of scientific figures based on paper content.
However, most conventional methods output the generated results as raster graphics, making post-editing (e.g., text modification and layout changes) highly difficult.
This poses a significant challenge, as they are unsuitable for the iterative figure revision process inherent in paper writing and peer review.
To tackle these challenges, we define the novel task of generating editable GAs from paper content and propose GenGA, a new GA generation framework that directly produces figures in vector format.
By generating figures as a collection of vector elements with a hierarchical structure, GenGA produces outputs that can be seamlessly imported into existing drawing tools for intuitive, element-level editing.
Furthermore, we introduce the Structural Independence Coefficient (SIC), a metric that quantifies the editing simplicity of a figure based on the degree to which local modifications propagate to other elements.
Experimental results show that GenGA achieves superior editing simplicity compared to conventional methods, and even surpasses human-authored GAs in conciseness and semantic alignment.
We also validate SIC as an effective metric correlated with manual editing costs.
This study fundamentally redefines GA generation as an editable vector graphic generation problem grounded in the practical workflows of researchers, significantly promoting effective scientific communication.

\end{abstract}

%% file: fig/tex/01_teaser.tex
\begin{teaserfigure}
    \includegraphics[width=\textwidth]{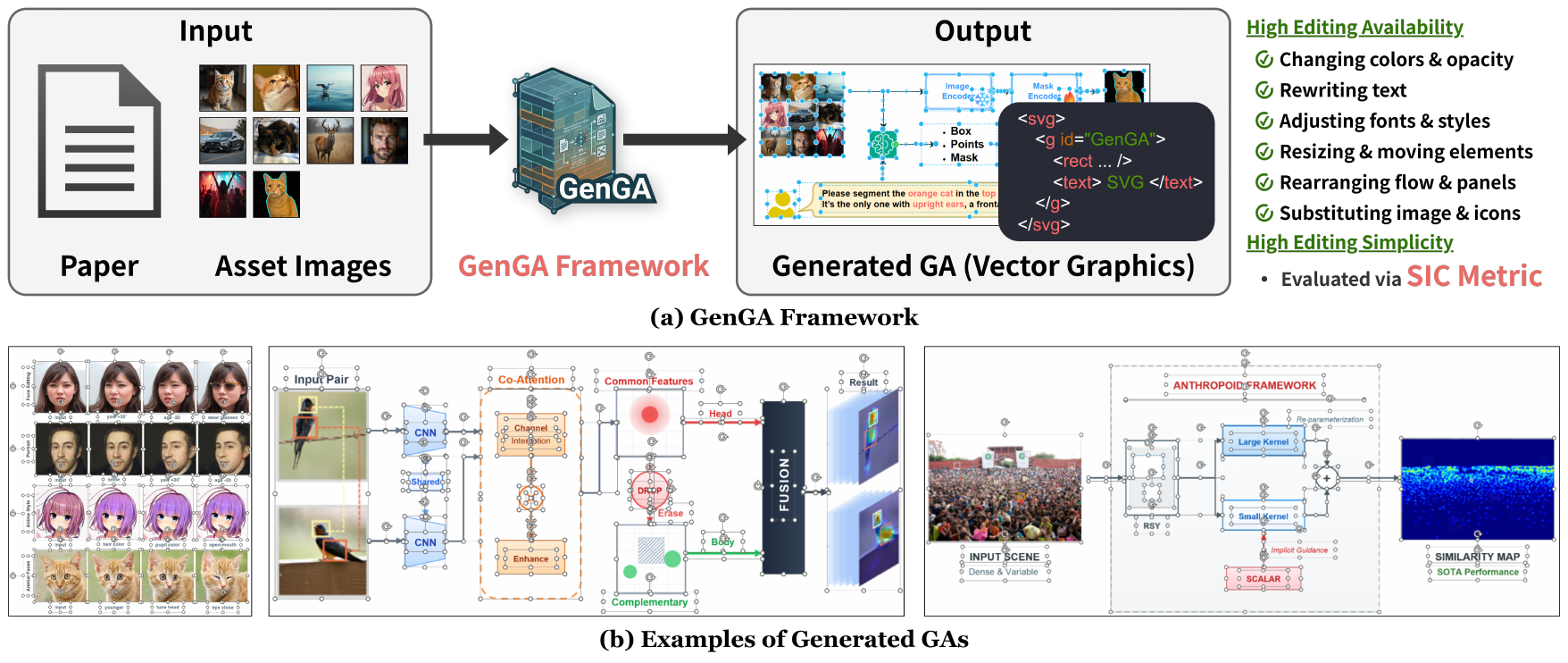}
    \vspace{-8mm}
    \caption{
        We introduce the task of Editable GA Generation, along with GenGA, a framework for generating editable GAs in vector format, and the Structural Independence Coefficient (SIC), a metric for quantifying editing simplicity.
        (a) The GenGA framework takes a paper and user-provided asset images as input and generates a GA in an editable vector format (SVG). The output supports flexible element-level editing, including layout adjustment, text modification, and image replacement, while maintaining high editing simplicity as measured by SIC.
        (b) Examples of generated GAs, illustrating structured layouts and element-wise editability in complex scientific visualizations.
        (Source papers:
        \href{https://arxiv.org/abs/2109.10737}{10.48550/arXiv.2109.10737},
        \href{https://arxiv.org/abs/2101.08527}{10.48550/arXiv.2101.08527},
        \href{https://arxiv.org/abs/2212.02248}{10.48550/arXiv.2212.02248})
    }
    \label{fig:teaser}
\end{teaserfigure}

%% file: sec/01_introduction.tex
\section{Introduction}

Scientific progress is driven by a continuous cycle of discovery and communication, yet researchers face constraints in time, expertise, and computational resources.
The automation of the discovery process has long been a central goal in scientific research~\cite{lenat1977math,lenat1983am,buchanan1978dendral}.
In recent years, under the paradigm of AI for Science, significant progress has been made toward automating scientific discovery, including hypothesis generation~\cite{merchant2023scaling, pyzer-Knapps2022accelerating, meincke2023llm-idea-gen} and experimental design~\cite{szymanski2023auto-lab, baek2025research-agent}.
In contrast, the process of effectively communicating research outcomes has not been automated to a comparable extent.
While AI-assisted tools for paper writing~\cite{wen2024overleafcopilot, lu2024ai-scientist} and presentation material generation~\cite{fu2022doc2ppt, pang2025paper2poster} have improved the efficiency of scientific communication, organizing complex research ideas into intuitive and visually understandable representations remains a non-trivial challenge.

Graphical Abstracts (GAs) are visual summaries that concisely present the key findings of research papers, often appearing as teasers or as \textit{Figure 1} at the beginning of papers.
GAs enable readers to quickly grasp the overall contribution of a study and enhance its visibility and impact~\cite{ibrahim2017va, kim2022Seeing, bennett2023ga}.
Despite the growing adoption of GAs, methods for effectively designing such visual content remain underdeveloped.
Designing a compelling GA requires careful prioritization of information and thoughtful visual composition~\cite{lee2023ga, jeyaraman2023attract}, posing a challenge for many researchers.
The generation of GAs and related scientific visualizations has recently attracted increasing attention.
For such visual content to be practically useful, beyond alignment with the paper content and visual quality, two additional requirements are critical.
First, the generated results must be easily editable.
In real research workflows, figures are not final products but are repeatedly revised during writing and peer review.
Changes in layout, emphasis, and content, such as adding or removing elements, are routine, making post-hoc manual editing a critical requirement.
Second, the method must reliably incorporate researcher-provided data.
Scientific figures often include selected examples, experimental results, and visualizations based on real data.
Automatically generating or synthesizing such content risks introducing hallucinated or incorrect information.
Therefore, concrete data and images should either be directly embedded or remain editable for manual insertion and refinement by researchers.

Recent advances in image generation models, such as GPT-Image-1 \cite{gpt-image} and NanoBanana-Pro~\cite{nanobanana}, have enabled the generation of visually appealing scientific figures at a practical level.
However, most conventional approaches~\cite{rodriguez2023fig-gen, zhu2026paperbanana, zhu2026autofigure, huang2026scifig} treat generated figures as final outputs, without considering structural editability for human post-editing or the reliable integration of real data.
Many methods~\cite{rodriguez2023fig-gen, zhu2026paperbanana} generate results as raster graphics.
Since raster graphics are merely collections of pixels and do not preserve structural information, they are difficult to edit after generation.
AutoFigure~\cite{zhu2026autofigure} partially addresses this limitation by extracting text and icons from generated raster graphics and reconstructing them as vector elements.
However, most of the figure remains as a background raster, making it impossible to edit the overall layout structure.
Furthermore, these methods do not assume the structural integration of user-provided data and instead attempt to generate all visual elements.
As a result, challenges remain in producing reliable scientific figures grounded in real data.

To address these limitations, we introduce a new task termed \textit{Editable GA Generation}.
Instead of generating GAs as static raster graphics, this task aims to produce editable vector graphics that support human post-editing.
In particular, we emphasize two key requirements: the editability of the generated results and the reliable integration of researcher-provided data.
To realize this task, we propose two core components.
First, we introduce the \textit{Structural Independence Coefficient (SIC)}, a metric that quantifies editing simplicity based on the structural independence of visual elements.
It evaluates editing locality, ensuring that local edits remain local without global side effects.
A high SIC reflects a highly editable figure where local edits remain local without global side effects.
Second, we propose \textit{GenGA}, a framework illustrated in Figure~\ref{fig:teaser} that directly generates editable vector GAs in SVG format.
By explicitly representing structural information, GenGA enables element-level editing while preserving user-provided data.
The generated outputs can be readily imported into common drawing tools such as PowerPoint, Adobe Illustrator, Figma, and draw.io, allowing seamless manual refinement within existing workflows.

Our main contributions are summarized as follows:
\begin{itemize}
    \item We define a new task, Editable GA Generation, which reformulates GA generation from a static image synthesis problem into a structural generation problem that explicitly supports iterative human editing.
    \item We introduce the SIC, a metric that formalizes editability as structural independence and enables quantitative evaluation of editing simplicity.
    \item We introduce GenGA, a vector-first and data-grounded framework that generates editable GAs while supporting seamless post-editing and integration of researcher-provided data.
\end{itemize}

%% file: sec/02_related_work.tex
\section{Related Work}

\paragraph{The Importance of GA and its Design.}
GAs are an effective means of visually communicating the key findings of scientific papers and have been shown to increase online visibility and engagement~\cite{ibrahim2017va, huang2018the-effect, chapman2019randomizad, hoffberg2020beyond, kunze2021infographics, kim2022Seeing, bennett2023ga}.
At the same time, overly abstract visual summaries may lead to misinterpretation, highlighting the difficulty of designing effective GAs~\cite{jeyaraman2023ga, jeyaraman2023attract}.
Prior work has explored design principles and reusable patterns for scientific figures and GAs~\cite{millar2022the-role, jeyaraman2023ga, jeyaraman2023attract, yuanyuan2023ga}.
More recently, \textcite{kawada2025sciga} established a data-driven foundation for GA research by constructing a large-scale dataset SciGA-145k and introducing GA recommendation tasks.
Furthermore, the direction of automated GA generation has been explicitly suggested in recent studies~\cite{kawada2025sciga, lee2023ga, kirukowski2023potential}.

\paragraph{Scientific Figure Generation.}
Recent advances in image generation models, such as latent diffusion models~\cite{rombach2022latent-diffusion-model} and NanoBanana-Pro~\cite{nanobanana}, have significantly improved the generation of visually compelling scientific figures from paper content~\cite{rodriguez2023fig-gen, zhu2026paperbanana, zhu2026autofigure, zhu2026autofigure-edit, huang2026scifig, wang2025scisketch}.
While these approaches achieve strong visual quality, they either fall short in producing aesthetically well-structured GAs or hinder post-editing, or do not consider editability.
Some approaches, such as AutoFigure~\cite{zhu2026autofigure}, introduce partial structure by overlaying vectorized text and icon elements on a raster background, where only limited foreground elements are editable while the overall layout remains fixed.
In addition to structural limitations, another important issue lies in how visual content is generated.
Most conventional approaches synthesize all visual elements, including experimental results and example images, through pixel generation, which can unintentionally alter or distort the underlying data, raising concerns about reliability in scientific contexts.
In contrast, we introduce a task centered on editability and propose an editable vector-first, data-grounded GA generation framework.

\paragraph{Editable and Vector-based Generation.}
Work on generating visual representations in vector or code form (e.g., SVG and TikZ) has enabled structured outputs for diagrams.
SVG-based models~\cite{jain2023vectorfusion, rodriguez2025starvector, yang2025omnisvg} produce outputs that can be easily imported into drawing tools, but are limited to simple designs such as icons or logos.
TikZ-based models~\cite{belouadi2024detikzify, belouadi2024automatikz, wei2025diagram-agent} target diagram generation but remain limited in scope and do not capture the complexity of GAs.
As a result, conventional methods do not adequately support the design of GAs, which require both coherent visual composition and structured representation of scientific content.

%% file: sec/03_proposed.tex
\section{Proposed Tasks, Metric, and Framework}

In this paper, we consider the practical workflow of GA creation, where editable and reliable figures grounded in real data are required, and make the following three contributions:
(1) We formulate Editable GA Generation as a structural generation problem that produces editable vector graphics.
(2) We introduce the Structural Independence Coefficient (SIC), a metric for evaluating the editing simplicity of generated GAs.
(3) We introduce GenGA, a framework for solving Editable GA Generation.
Details of each component are provided in the following subsections.

\subsection{Task Formulation}
We consider the task of generating GAs that enable post-editing.
We define editability as the ability to modify a figure through element-level operations (e.g., modifying text content, changing shape colors, adding elements, adjusting arrow connections, modifying layouts, and replacing images), characterized by two complementary aspects.
First, \textit{editing availability} refers to whether a target operation can be performed under a given representation.
This depends on the compatibility between the representation and the editing task.
For example, raster graphics generally do not support operations such as text rewriting, whereas representations with explicit structural elements enable such operations.
Second, \textit{editing simplicity} refers to the amount of effort required to perform an available operation.
An edit is considered simple if it can be completed by modifying a single element, whereas it becomes more complex when it requires coordinated changes across multiple elements.

Based on these criteria, we define Editable GA Generation as the task of generating a representation that satisfies both editing availability and simplicity, which is formulated as:
\begin{equation}
\Phi : (T_{\text{full-text}}, A) \rightarrow I_{\text{vector}},
\end{equation}
where $T_{\text{full-text}}$ is the full text of a target paper, $A$ is a set of user-provided assets, and $I_{\text{vector}}$ is a vector-based GA.
The asset set is defined as $A = \{ (a_{\text{raster}}^{(m)}, a_{\text{caption}}^{(m)}) \mid m \in \{1, 2, \dots, N_\mathrm{a}\} \}$, where $a_{\text{raster}}^{(m)}$ denotes a raster graphic (e.g., an input-output example or a device photograph), and $a_{\text{caption}}^{(m)}$ denotes its corresponding caption.
Each $a_{\text{raster}}^{(m)}$ represents a visual component that is incorporated into the final GA.
The output is represented as $I_{\text{vector}} = \{ z_i \mid i \in \{1, 2, \dots, N_\mathrm{z} \} \}$, where each $z_i$ denotes an individual graphical element, enabling element-level editing in standard drawing tools.
Each element $z_i$ is defined as a structured object
$z_i = (\tau_i, x_i, y_i, w_i, h_i, \gamma_i, \pi_i, \rho_i)$,
where $\tau_i$ denotes the element type (e.g., \texttt{<text>},\allowbreak \texttt{<rectangle>},\allowbreak \texttt{<circle>},\allowbreak \texttt{<line>},\allowbreak \texttt{<arrow>} or \texttt{<image>}),
$(x_i, y_i) \in \mathbb{R}^2$ represents its position in the canvas coordinate system, and $(w_i, h_i) \in \mathbb{R}^2$ denotes its size.
$\gamma_i$ represents visual attributes such as stroke color, fill color, and opacity, and $\pi_i \in \{1, \dots, N_\mathrm{z}\} \cup \{\varnothing\}$ denotes the index of the parent element, defining the hierarchical structure.
$\rho_i \in \{\texttt{exclusive}, \texttt{overlay}\}$ specifies whether elements at the same hierarchical level are allowed to overlap.
In particular, elements with $\rho_i = \texttt{exclusive}$ must not spatially overlap.
Each user-provided asset $a_{\text{raster}}^{(m)}$ is incorporated as an element with $\tau_i = \texttt{<image>}$.

\subsection{Structural Independence Coefficient (SIC)}
We next introduce the Structural Independence Coefficient (SIC), a metric for evaluating the editing simplicity of a figure that can be applied to both vector and raster graphics.
SIC models a figure as a dependency graph over its elements and quantifies the extent to which a local modification to one element propagates to others, yielding a score in $[0, 1]$.
A higher SIC indicates that the effects of edits remain localized, allowing elements to be modified more independently.

\paragraph{Definition.}
We model a figure $I$ as a weighted complete undirected graph $G = (V, E)$ that represents dependencies among visual elements.
Here, $V = \{ v_j \mid j \in \{1, 2, \dots, N_\mathrm{v} \} \}$ denotes the set of nodes, where each $v_j$ corresponds to a visual component in the figure (e.g., text, shapes, arrows, or images).
$E = \{ (v_j, v_k, w_{jk}) \mid j < k \}$ denotes the set of weighted edges.
Each weight $w_{jk} \in [0,1]$ quantifies the likelihood that a local edit to $v_j$ propagates to $v_k$, where larger values indicate stronger mutual dependency between the elements.
By construction, we have $w_{jk} = w_{kj}$.
We further assume that edit effects propagate transitively along dependencies between visual elements.
To model this stochastic propagation process, we perform independent Bernoulli trials on each edge $(v_j, v_k)$, activating the edge with probability $w_{jk}$.
This yields a random edge set $\tilde{E} \subseteq E$ and the corresponding random graph $\tilde{G} = (V, \tilde{E})$.
The number of elements affected by editing $v_j$ is given by the size of the connected component containing $v_j$ in $\tilde{G}$, denoted as $n_j$.
Using this, we define the Structural Independence Coefficient (SIC) of a figure $I$ as:
\begin{equation}
\mathrm{SIC}(I) = 1 - \mathbb{E}_{\tilde{G}} \left[ \frac{1}{N_\mathrm{v}} \sum_{v_j \in V} \frac{n_j-1}{N_\mathrm{v}-1} \right].
\end{equation}
Here, $(n_j-1) / (N_\mathrm{v} - 1)$ represents the proportion of elements that may be affected when editing $v_j$.
Therefore, SIC can be interpreted as the complement of the expected edit propagation over the figure, taking values in $[0, 1]$, where higher values indicate that edits remain localized, allowing elements to be modified more independently.

\input{fig/tex/02_GenGA}
\input{tbl/01_comparison}

\paragraph{Instantiating nodes $V$ and edges $E$.}
To compute SIC in practice, we instantiate the abstract graph $G = (V, E)$ from a given figure $I$ by defining its editing units $V$ and edge weights $\{w_{jk}\}$ according to the representation of $I$, i.e., whether the figure is represented as vector graphics or raster graphics.
(i) In vector graphics, visual elements such as text, shapes, and arrows are explicitly represented as structured objects.
Each element is treated as an editing unit and mapped to a node in $V$.
Edge weights are defined based on explicit structural relationships.
Specifically, if two elements $(v_j, v_k)$ share structural constraints (e.g., grouping, common transformations, or clipping), then $w_{jk}=1$; otherwise, $w_{jk}=0$.
Raster graphics embedded in vector graphics are normally treated as atomic \texttt{<image>} nodes.
As an exception, when such an image dominates the figure, we regard the entire figure as a raster-dominant representation and apply the raster formulation described below.
(ii) In raster graphics, explicit structural elements are not available.
We therefore approximate editing units as visually homogeneous regions obtained via Felzenszwalb segmentation~\cite{felzenszwalb2004efficient}, and treat each region as a node in $V$.
This is motivated by the observation that, in common painting tools, operations such as recoloring or background modification can be performed at the region level.
Edge weights are defined as:
\begin{equation}
w_{jk} = \exp(-b_{jk}),
\end{equation}
where $b_{jk}$ denotes the boundary strength between adjacent regions.
The boundary strength is computed as:
\begin{equation}
b_{jk} = \frac{1}{|B_{jk}|} \sum_{p \in B_{jk}} \frac{||\nabla I(p)||}{g_{\max}},
\end{equation}
where $B_{jk}$ is the set of boundary pixels, $\nabla I(p)$ is the image gradient at pixel $p$ computed using the Sobel operator, and $g_{\max} = 4\sqrt{2}$ is the theoretical maximum gradient magnitude.
This formulation ensures that strong boundaries result in weak dependencies, thereby limiting edit propagation across regions.

\subsection{GenGA Framework}
We introduce GenGA, a framework for solving Editable GA Generation.
GenGA takes as input a paper $T_{\text{full-text}}$ and optionally assets $A$ (e.g., an input-output example or a device photograph), and generates an editable GA in SVG format $I_\mathrm{vector}$, preserving editing availability that conventional methods lack (see Table~\ref{tbl:comparison}).
As illustrated in Figure~\ref{fig:GenGA}, the framework consists of the following four phases:
(1) Reference Retrieval:
Given the input paper $T_{\text{full-text}}$, we retrieve GAs from existing papers that are semantically relevant and use them as visual references.
(2) Reference Vectorization:
The retrieved raster-format references are converted into structured vector representations, capturing layout and structural information while abstracting away non-structural content.
(3) Asset-aware Generation:
Given $T_{\text{full-text}}$, assets $A$, and the references, we generate a GA in SVG format.
(4) Self-correction Loop:
The generated result is iteratively reviewed and refined to improve its quality.

\paragraph{Reference Retrieval.}
This phase retrieves a semantically relevant GA as a reference.
Using a text-to-image retrieval model $f_{\text{retrieve}}(\cdot)$, we retrieve a reference GA as:
\begin{equation}
I_{\text{ref}}^{\text{raster}} = f_{\text{retrieve}}(T_{\text{full-text}}, R_{\text{raster}}),
\end{equation}
where $T_{\text{full-text}}$ is the input paper and $R_{\text{raster}} = \{ r_\ell \mid \ell \in \{1,2,\dots,N_r\} \}$ denote a set of candidate raster-format GAs.
In the subsequent asset-aware generation phase, the retrieved reference provides domain-specific visual conventions.

\paragraph{Reference Vectorization.}
This phase converts the retrieved raster-format references into a structured vector representation.
This conversion is performed using a vision-language model (VLM) $f_{\text{vectorize}}(\cdot)$ as follows:
\begin{equation}
I_{\text{ref}}^{\text{vector}} = f_{\text{vectorize}}\left(I_{\text{ref}}^{\text{raster}}\right).
\end{equation}
The vectorization model extracts structural components such as regions, arrows, and text blocks, and represents them as elements in SVG format.
Importantly, non-structural elements such as photographs, icons, plots, and example inputs are replaced with simple placeholder elements (i.e., rectangular regions) that preserve only position and size without encoding visual content.
This design avoids introducing unreliable geometric approximations of non-structural content, e.g., approximating photographic content using geometric primitives such as lines and shapes.
In the subsequent asset-aware generation phase, the vectorized reference provides structural guidance for generating GAs in SVG format.

\paragraph{Asset-aware Generation.}
This phase generates a GA as a structured SVG while explicitly incorporating user-provided assets as non-generative constraints.
An initial vector representation is generated using a VLM $f_{\text{generate}}(\cdot)$:
\begin{equation}
I_{\text{vector}}^{(0)} = f_{\text{generate}}\left(T_{\text{full-text}}, A, I_{\text{ref}}^{\text{raster}}, I_{\text{ref}}^{\text{vector}}\right).
\end{equation}
Crucially, user-provided assets are not treated as generative targets but as fixed structural constraints, and are represented as placeholder elements.
Unlike prior approaches that generate all visual content end-to-end, GenGA treats scientific data as immutable inputs rather than generative targets, ensuring that experimentally derived content is neither hallucinated nor altered during generation.
As a result, the generated GA remains faithful to the original data while maintaining structural consistency and editability, enabling reliable and data-grounded scientific visualization.

\paragraph{Self-correction Loop.}
This phase iteratively improves the generated GA through feedback-driven refinement.
Starting from the initial result $I_{\text{vector}}^{(0)}$, the model performs iterative review and refinement.
Let $I_{\text{vector}}^{(t)}$ denote the vector representation at iteration $t$.
Since SVG is a code-based representation, visual evaluation is performed on its rasterized form:
\begin{equation}
I_{\text{raster}}^{(t)} = f_{\text{rasterize}}\left(I_{\text{vector}}^{(t)}\right).
\end{equation}
Feedback is obtained via a VLM $f_{\text{review}}(\cdot)$:
\begin{equation}
T_{\text{feedback}}^{(t)} = f_{\text{review}}\left(T_{\text{full-text}}, I_{\text{raster}}^{(t)}\right).
\end{equation}
The reviewer evaluates aspects such as semantic consistency, layout conflicts, and readability.
The generator $f_{\text{generate}}(\cdot)$ then regenerates the output using both $I_{\text{vector}}^{(t)}$ and $T_{\text{feedback}}^{(t)}$:
\begin{equation}
I_{\text{vector}}^{(t+1)} = f_{\text{generate}}\left(T_{\text{full-text}}, A, I_{\text{vector}}^{(t)}, T_{\text{feedback}}^{(t)}\right).
\end{equation}
This process is repeated for $t = 0, \dots, N_t - 1$.

%% file: fig/tex/02_GenGA.tex
\begin{figure*}[!t]
    \centering
    \includegraphics[width=\linewidth]{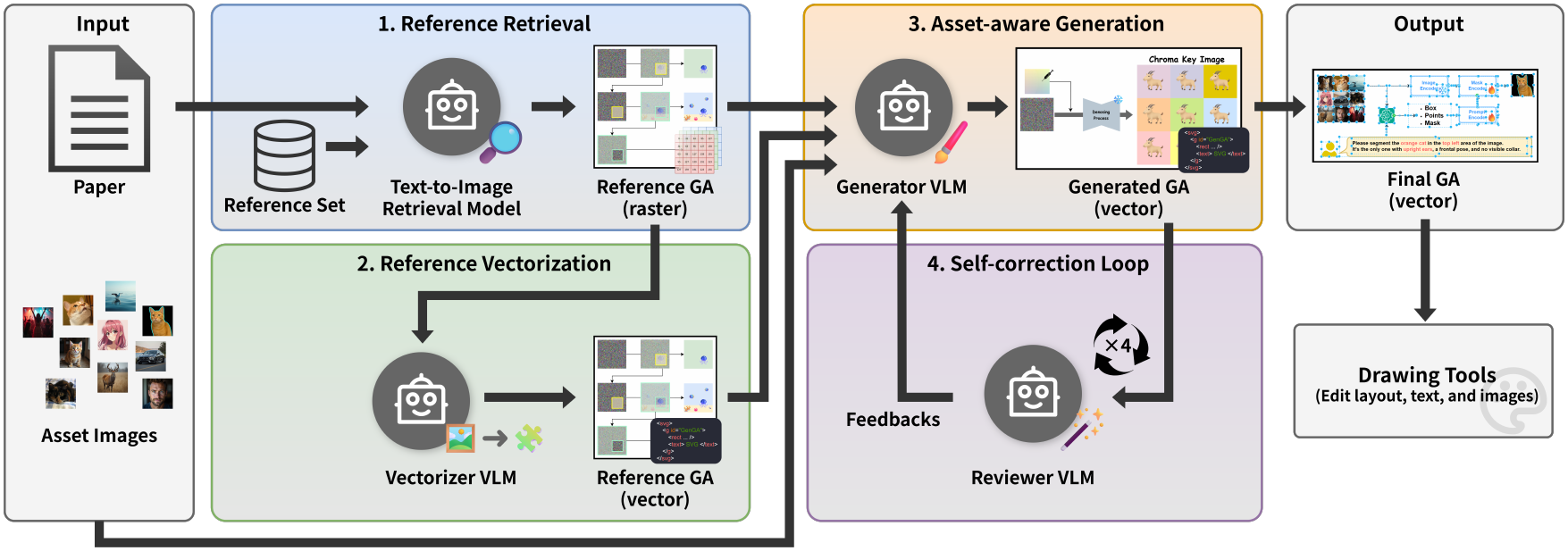}
    \caption{
        Overview of the GenGA framework.
        Given a paper and optional asset images, GenGA generates an editable GA in SVG format, enabling flexible element-level editing and seamless integration into existing tools.
        The model retrieves and vectorizes reference GAs to capture visual and structural patterns, and generates the output accordingly.
        A self-correction loop further refines the result to improve semantic alignment, layout, and readability.
    }
    \label{fig:GenGA}
\end{figure*}

%% file: tbl/01_comparison.tex
\begin{table}[t]
\centering
\caption{
Comparison of post-editing capabilities across figure generation methods.
GenGA uniquely enables fine-grained, localized edits while preserving structural consistency, highlighting its advantage in controllable and reusable figure generation.
}
\label{tbl:comparison}
\begin{tabular}{@{}lccc@{}}
\toprule
\textbf{Post-Editing Task} & \begin{tabular}[c]{@{}c@{}}\textbf{Paper}\\ \textbf{Banana}\\ \cite{zhu2026paperbanana}\end{tabular} & \begin{tabular}[c]{@{}c@{}}\textbf{Auto}\\ \textbf{Figure}\\ \cite{zhu2026autofigure}\end{tabular} & \textbf{\begin{tabular}[c]{@{}c@{}}GenGA \\ (ours)\end{tabular}} \\ \midrule
Change colors \& opacity       & \cmark                                                                             & \cmark                                                                           & \cmark                                                           \\
Rewrite text                   & \xmark                                                                             & \cmark                                                                           & \cmark                                                           \\
Adjust fonts                   & \xmark                                                                             & \cmark                                                                           & \cmark                                                           \\
Replace images \& icons        & \xmark                                                                             & \cmark                                                                           & \cmark                                                           \\
Resize \& move elements        & \xmark                                                                             & \xmark                                                                           & \cmark                                                           \\
Rearrange flow \& panels       & \xmark                                                                             & \xmark                                                                           & \cmark                                                           \\ \bottomrule
\end{tabular}
\end{table}

%% file: sec/04_experiments.tex
\section{Experimental Setup}
We conduct experiments on generating GAs from the abstract and introduction of papers in markdown format.
We use the large-scale SciGA-145k dataset~\cite{kawada2025sciga}, which contains papers paired with GAs, and focus on the Computer Science domain, which has the largest number of samples.
For evaluation, we use a test set of 2,053 papers.
In the reference retrieval phase, we use an additional set of 16,416 GAs, disjoint from the evaluation set, as retrieval candidates.

Assets are extracted from the original GAs using a semi-automatic pipeline.
We apply a layer decomposition model, LayerD~\cite{suzuki2025layerd}, to obtain candidate regions, retaining raster content while excluding structural elements such as text and shapes.
Each region is assigned a semantic caption describing its role and content within the figure using Gemini-3.1-Pro-Preview~\cite{gemini3}.
All extracted regions and captions are then formally verified by the authors following a predefined guideline, and curated through discussion in cases of clear mismatch.

\subsection{Implementation Details}
We use GPT-5.2~\cite{gpt5} and Gemini-3.1-Pro-Preview as the backbone models for the generator and reviewer, respectively, and compare their performance.
In the reference retrieval phase, we employ Long-CLIP-4-Inter-GA-Rec~\cite{kawada2025sciga} as the text-to-image retrieval model and Gemini-3.1-Pro-Preview as the vectorization model.
Long-CLIP-4-Inter-GA-Rec retrieves existing GAs based on their similarity to the input abstract, originally designed to support human inspiration during GA creation.
In this work, we extend its use beyond human assistance and leverage it as a structural retrieval module that provides guidance for machine-driven generation.
In the self-correction loop phase, we perform four iterations of refinement.

\subsection{Baseline Methods}
We compare GenGA with the following three methods:

\textit{(i) Author-created GAs.}
GAs created by human authors are used as an upper bound, representing ideal readability and faithfulness.

\textit{(ii) Raster-based generation methods.}
This category, including NanoBanana-Pro~\cite{nanobanana} and PaperBanana~\cite{zhu2026paperbanana}, which directly generate scientific figures as raster graphics.
While these methods achieve high visual quality, their outputs are pixel-locked and lack structural editability.
To provide a stronger baseline, we additionally convert their outputs into vector format using Gemini-3.1-Pro-Preview.
This setting represents a simple extension of raster-based generation toward editability, allowing us to examine whether post-hoc vectorization can recover structural manipulability.

\textit{(iii) Hybrid raster-vector methods.}
We include AutoFigure~\cite{zhu2026autofigure}, which maintains the background as a single raster graphic while overlaying text and icons as independent vector elements.
Although this design enables partial editability, the overall structure remains largely fixed as a single raster component, limiting structural modifications.

\subsection{Evaluation Metrics}
We evaluate Editable GA Generation from multiple perspectives using the following five complementary metrics:

\textit{(i) CLIP-S (Semantic Alignment).}
We measure the text--image CLIPScore~\cite{hessel2020clipscore} between the paper abstract and the generated GA. 
Since a GA serves as a visual summary of a paper, the abstract is used as a proxy for the underlying content.

\textit{(ii) Overlap Ratio (Layout Quality).}
We assess geometric clarity and layout quality using the Overlap Ratio, defined as the ratio of overlapping areas between bounding boxes of SVG elements to the total area. 
Overlaps between parent--child elements are excluded to avoid penalizing hierarchical structures.

\textit{(iii) SIC (Editability).}
We evaluate editability using the proposed SIC, which measures the extent to which local edits can be performed without affecting other elements.

\textit{(iv) Contribution QA Accuracy (Content Understanding).}
We evaluate how well the generated GA conveys the paper’s main contribution using the accuracy of a visual question answering, where a VLM selects the most relevant statement given the generated GA.
We construct the task by generating a ground-truth contribution statement from the full text using GPT-5.2, and sampling three hard negative candidates from contribution statements of other papers whose abstracts are most similar under Sentence-BERT~\cite{reimers2019sentence-bert}, forming a four-choice question.
Both question generation and answering are performed by GPT-5.2 and Gemini-3.1-Pro-Preview.

\textit{(v) VLM-as-a-Judge (Subjective Evaluation).}
We adopt a VLM-based evaluation protocol following prior work~\cite{zhu2026paperbanana}, where GPT-5.2 and Gemini-3.1-Pro-Preview are used as evaluators.
Generated GAs are evaluated along four dimensions: Faithfulness (consistency with the paper content), Conciseness (ability to summarize key contributions without redundancy), Readability (clarity of layout and relationships), and Aesthetics (visual quality appropriate for academic figures). 
The VLM compares generated GAs with human-authored GAs and assigns scores of 1.0 (win), 0.5 (tie), or 0.0 (loss). 
For overall evaluation, we employ a hierarchical strategy in which Faithfulness and Conciseness are treated as primary criteria, and Readability and Aesthetics as secondary criteria. 
If a clear winner is determined by the primary criteria, that result is adopted; otherwise, the decision is based on the secondary criteria.

\subsection{User Study}
\paragraph{Manual Editing Cost.}
To evaluate the actual editing cost of figures and validate the effectiveness of SIC as its proxy, we conduct a user study with 15 professional machine learning researchers who have experience in GA creation.
Participants are given visually identical figures that differ in SIC due to representation and structural complexity, and are asked to edit multiple predefined targets (8 per figure), including text modification, visual adjustments, and image replacement, to match a target within a 5-minute time budget.
Editing is performed using draw.io, and interaction logs are recorded.
From these logs, we measure editing time treated as right-censored at the time limit, and the fraction of successful edits.

\paragraph{Human Preference.}
To evaluate the subjective quality and practical usability of generated GAs, we conduct a two-alternative forced choice (2AFC) user study.
For each query paper, participants are presented with a pair of GAs generated by different methods and asked to compare them.
Each participant evaluates a subset of the query papers, resulting in a total of 300 pairwise comparisons.
Participants are asked to select which GA they prefer overall, based on their impression considering factors such as visual quality, readability, and appropriateness as a scientific figure.
We report win rates, Bradley--Terry scores~\cite{bradley1952rank} and Elo ratings~\cite{elo1978rating}.
In addition, for each GA, participants independently indicate whether it is of sufficient quality to be directly published in a scientific paper.

%% file: sec/05_results_and_discussion.tex
\input{tbl/02_results}
\input{fig/tex/03_user_study}
\input{tbl/03_user_study}

\section{Results and Discussion}

\paragraph{SIC and Human Editing Performance.}
Figure~\ref{fig:user_study} summarizes the relationship between SIC and human editing performance.
Higher SIC consistently leads to a higher fraction of successful edits within the time limit and lower editing time, with strong correlations observed for each metric (respectively: $r = 0.950$, $p < 10^{-7}$ and $r = -0.967$, $p < 10^{-8}$).
We observe that users struggle with low-SIC figures due to strong interdependencies, which require cascading adjustments even for simple edits.
This occurs in tightly coupled layouts, grouped elements, and ambiguous or overlapping regions that hinder localized modifications.
For raster-based figures, participants sometimes resort to manually recreating elements from scratch and overlaying them on the original figure, where ambiguous boundaries further increase the required effort by forcing larger regions to be rebuilt.
Overall, these results show that SIC effectively captures practical editing cost in real-world figure revision.

\input{fig/tex/04_results}

\paragraph{Quantitative Results.}
Table~\ref{tbl:results} summarizes the quantitative evaluation results of GAs generated by each method.
Raster-based PaperBanana achieved the highest  Aesthetics scores (GPT-5.2: 0.692, Gemini-3.1-Pro-Preview: 0.455), confirming its strong rendering capability.
However, it exhibits low editability, as indicated by a low SIC score of 0.184.
Applying post-hoc vectorization improves SIC to 0.794 but consistently degrades other metrics. This reflects the ill-posed nature of inferring structure (e.g., grouping and hierarchy) from pixel-based representations, leading to information loss and geometric inconsistencies that hinder both visual quality and editability.
AutoFigure achieves a perfect Overlap Ratio of 0.000 due to its design, which fixes most of the figure as a single raster background and overlays only a limited number of vector elements such as text and icons.
While this avoids spatial interference between elements, the underlying structure remains largely uneditable, resulting in a low SIC score of 0.131.

In contrast, GenGA directly generates all elements as a hierarchical vector structure, achieving a strong balance between visual quality and editability.
In particular, GenGA with Gemini-3.1-Pro-Preview achieves a SIC score of 0.943, significantly outperforming all baselines.
Although its Overlap Ratio (0.163) appears higher, this reflects intentional layout design (e.g., text placed within structured regions) rather than undesirable overlaps, and is consistent with maintaining structural coherence.
GenGA also shows strong semantic alignment with the source paper.
It surpasses human-authored GAs in CLIP-S (0.284 vs. 0.255), indicating effective content representation.
Furthermore, it achieves the highest Contribution QA accuracy (0.478), outperforming both human-authored GAs and conventional methods.
This suggests that directly grounding generation in textual reasoning enables GenGA to preserve key contributions without the information degradation introduced by intermediate image generation.
In terms of overall quality, GenGA maintains competitive Aesthetics (0.438) while achieving high Conciseness (0.690) and Readability (0.683).
Although it does not always outperform human-authored GAs in VLM-as-a-Judge evaluations, this gap is offset by its significantly higher editability, highlighting the importance of considering editability as a core evaluation dimension.
Across backbone models, GPT-5.2 performs better in semantic alignment (CLIP-S, Faithfulness), while Gemini-3.1-Pro-Preview excels in information structuring (Conciseness, Readability).
Overall, GenGA establishes a new paradigm for GA generation that is both editable and grounded in real data, without sacrificing visual quality.

\paragraph{User Preference.}
Table~\ref{tbl:user_study} summarizes the results of the 2AFC user study.
GenGA is strongly preferred over all baselines, achieving the highest win rate (0.767), Bradley--Terry score (1.189), and Elo rating (1728), indicating a clear and consistent user preference.
GenGA also achieves the highest publishable rate (0.667), suggesting that users not only prefer its outputs but also consider them suitable for direct use in scientific papers.
We observe high inter-user agreement in preference judgments (77.2\%; Krippendorff’s $\alpha = 0.523$, Fleiss’ $\kappa = 0.520$), while agreement on publishability is moderate (61.8\%; $\alpha = 0.235$, $\kappa = 0.233$), reflecting differences in user strictness.

\paragraph{Qualitative Analysis.}
Figure~\ref{fig:results} provides a qualitative comparison of generated GAs. Consistent with the quantitative results, existing methods exhibit several issues, including the excessive use of icons that is uncommon in scientific figures, truncated or incomplete text, misaligned elements, and the generation of fabricated input--output image that overflow or collide with surrounding components.
Moreover, as these outputs are generated as raster graphics, such errors are difficult to correct.
In contrast, GenGA produces clean and well-structured vector representations with clearly separated regions.
This vector structure improves visual clarity and enables localized modifications without affecting unrelated elements.
These observations are consistent with the improvements in SIC and user editing performance, further supporting the effectiveness of our approach.

\input{tbl/04_ablation}

\paragraph{Ablation Study.}
We conduct an ablation study to analyze the contribution of each component in GenGA, including the reference input format (raster and vector) and the self-correction loop (Table~\ref{tbl:ablation}).
Incorporating reference retrieval provides consistent improvements across multiple metrics.
Vector-based references yield higher CLIP-S, suggesting stronger semantic alignment with the input paper, while raster references improve visual richness, reflected in higher aesthetics scores.
Combining both further improves performance, highlighting their complementary roles.
The self-correction loop further enhances performance, particularly in VLM-as-a-Judge metrics such as conciseness, readability, and overall quality, indicating that iterative refinement aligns the generated structure with both semantic and visual criteria.
The full model, integrating both reference types and the self-correction loop, achieves the best overall performance, with the highest VLM-as-a-Judge (overall: 0.473) and Contribution QA accuracy (0.467).
These results indicate that each component contributes complementary benefits and that their integration is key to achieving both high editability and high-quality GA generation.

\input{fig/tex/05_ablation}

We further analyze the effect of the self-correction loop across refinement iterations (Figure~\ref{fig:ablation}).
Most metrics improve significantly from the first to the second iteration, indicating that early refinement plays a critical role in aligning the generated structure with the input content.
The initial iteration tends to produce overly simplified outputs with low faithfulness, indicating missing information.
Subsequent iterations progressively enrich the content.
Faithfulness and Contribution QA steadily increase, reflecting improved information coverage, while conciseness converges to a balanced level, indicating an appropriate trade-off between brevity and informativeness.
Later iterations focus on fine-grained refinements, including layout adjustments, visual styling, and textual improvements, resulting in clearer and more polished outputs.
Performance stabilizes after approximately four iterations, indicating convergence of the refinement process.
Overall, the self-correction loop functions as an effective iterative optimization mechanism that balances structural consistency, semantic alignment, and information completeness, ultimately producing high-quality GAs.

%% file: tbl/02_results.tex
\begin{table*}[!t]
    \caption{
    Quantitative comparison across methods.
    Our method achieves the highest editability (SIC) while maintaining strong semantic alignment, readability, and the ability to effectively convey key contributions.
        The best results for each metric are highlighted in \textbf{bold}.
    }
    \label{tbl:results}
    \centering
    \small
    \resizebox{\textwidth}{!}{
        \begin{tabular}{lcccccccccc}
        
        \toprule
        
        \multirow{2}{*}{\textbf{Method}} &
        \multirow{2}{*}{\begin{tabular}[c]{@{}c@{}} \textbf{Output}\\ \textbf{Format}\end{tabular}} &
        \multirow{2}{*}{\textbf{CLIP-S} ($\uparrow$) } &
        \multirow{2}{*}{\begin{tabular}[c]{@{}c@{}} \textbf{Overlap} \\ \textbf{Ratio}\end{tabular} ($\downarrow$) } &
        \multirow{2}{*}{\textbf{SIC} ($\uparrow$) } &
        \multirow{2}{*}{\begin{tabular}[c]{@{}c@{}} \textbf{Contribution QA } \\ \textbf{Accuracy}$^\dagger$ \end{tabular} ($\uparrow$)} &
        \multicolumn{5}{c}{\textbf{VLM-as-a-Judge}$^\dagger$} \\
        
        \cline{7-11}
        
        & & & & & &
        \multirow{1.25}{*}{\textbf{Faithfulness} ($\uparrow$)} &
        \multirow{1.25}{*}{\textbf{Conciseness} ($\uparrow$)} &
        \multirow{1.25}{*}{\textbf{Readability} ($\uparrow$)} &
        \multirow{1.25}{*}{\textbf{Aesthetics} ($\uparrow$)} &
        \multirow{1.25}{*}{\textbf{Overall} ($\uparrow$)} \\
        
        \midrule
        
        Author-created GA                     & Raster & 0.255 &          --    & 0.119 & 0.387 / 0.432 & 0.500 / 0.500 & 0.500 / 0.500 & 0.500 / 0.500 &        0.500 / 0.500  & 0.500 / 0.500 \\
        \midrule
        NanoBanana-Pro~\cite{nanobanana}          & Raster & 0.230 &          --    & 0.188 & 0.411 / 0.442 & 0.340 / 0.044 & 0.319 / 0.511 & 0.643 / 0.223 &        0.677 / 0.402 & 0.338 / 0.144 \\
        \quad + vectorize                     & Vector & 0.222 &         0.169  & 0.771 & 0.387 / 0.426 & 0.307 / 0.071 & 0.232 / 0.464 & 0.307 / 0.126 &        0.209 / 0.142 & 0.170 / 0.083 \\
        PaperBanana~\cite{zhu2026paperbanana} & Raster & 0.250 &          --    & 0.184 & 0.419 / 0.462 & 0.375 / 0.070 & 0.321 / 0.537 & 0.660 / 0.272 &\textbf{0.692 / 0.455}& 0.388 / 0.194 \\
        \quad + vectorize                     & Vector & 0.242 &         0.164  & 0.794 & 0.392 / 0.457 & 0.292 / 0.065 & 0.275 / 0.457 & 0.358 / 0.144 &        0.171 / 0.204 & 0.188 / 0.101 \\
        AutoFigure~\cite{zhu2026autofigure}   & Vector & 0.254 & \textbf{0.000} & 0.131 & 0.417 / 0.422 & 0.268 / 0.025 & 0.282 / 0.302 & 0.348 / 0.025 &        0.545 / 0.259 & 0.214 / 0.048 \\
        \midrule
        \rowcolor[HTML]{EFEFEF}
        GenGA (ours) & & & & & & & & & & \\
        \quad GPT-5.2~\cite{gpt5}                   & Vector &\textbf{0.284}& 0.146 &        0.885 &\textbf{0.427 / 0.478}&\textbf{0.531 / 0.274}&        0.263 / 0.612 &        0.672 / 0.288 & 0.654 / 0.318 &        0.389 / 0.315  \\
        \quad Gemini-3.1-Pro-Preview~\cite{gemini3} & Vector &        0.269 & 0.163 &\textbf{0.943}&        0.422 / 0.467 &        0.427 / 0.198 &\textbf{0.473 / 0.690}&\textbf{0.683 / 0.392}& 0.690 / 0.438 &\textbf{0.473 / 0.362} \\

        \bottomrule
        
        \end{tabular}
    }

\vspace{2pt}
\hfill {\footnotesize $\dagger$ Scores before/after the slash are reported using GPT-5.2 / Gemini-3.1-Pro-Preview, respectively, as the evaluator.}

\end{table*}

%% file: fig/tex/03_user_study.tex
\begin{figure}[!t]
    \centering
    \begin{minipage}{0.45\linewidth}
        \centering
        \includegraphics[width=\textwidth]{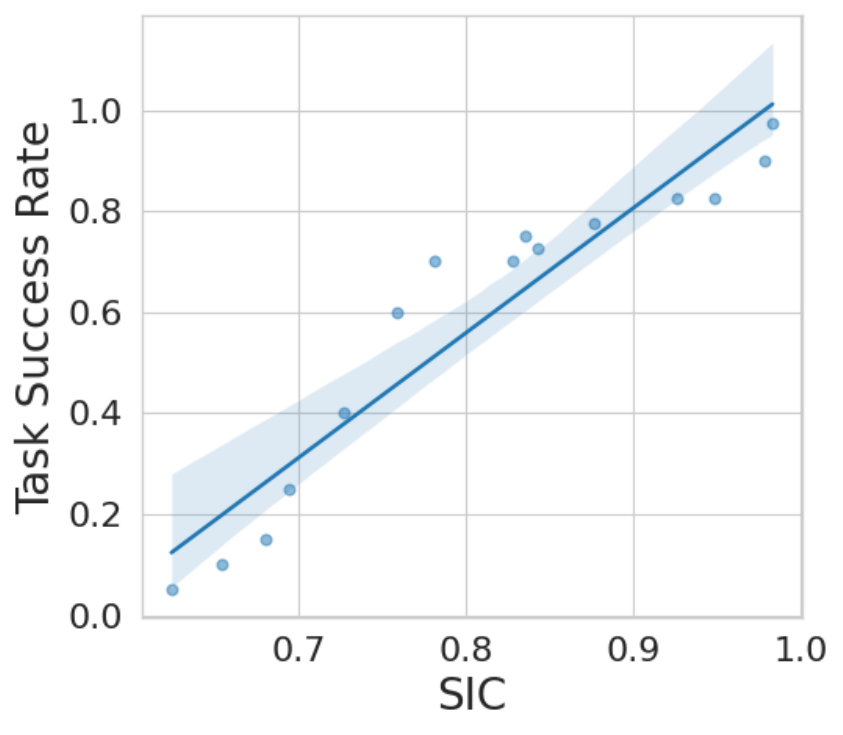}
        \subcaption{Task Success Rate}
        \label{fig:user_study:a}
    \end{minipage}   
    \begin{minipage}{0.45\linewidth}
        \centering
        \includegraphics[width=\textwidth]{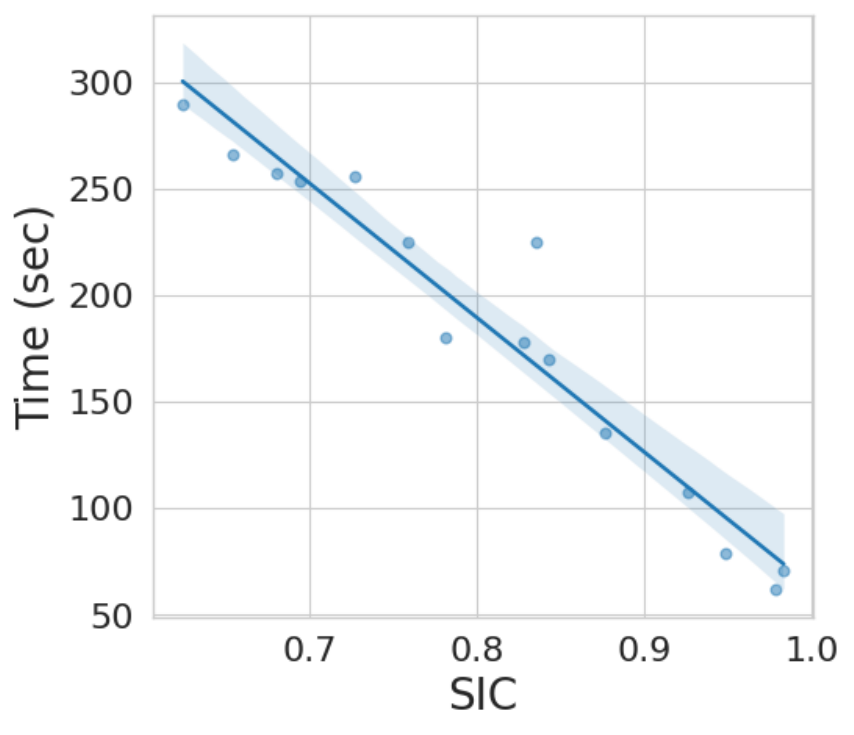}
        \subcaption{Editing Time}
        \label{fig:user_study:b}
    \end{minipage}
    \caption{
        Relationship between SIC and human editing performance.
        Higher SIC is associated with (a) a higher fraction of successful edits and (b) lower editing time, showing strong correlations with both metrics and demonstrating that SIC effectively captures practical editing cost.
    }
    \label{fig:user_study}
\end{figure}

%% file: tbl/03_user_study.tex
\begin{table}[!t]
    \caption{
        User preference evaluation.
        GenGA is consistently preferred over all baselines in pairwise comparisons and achieves the highest publishable rate, highlighting its superior perceived quality and practical usability.
        The best results for each metric are highlighted in \textbf{bold}.
    }
    \label{tbl:user_study}
    \centering
    \small
    \resizebox{\columnwidth}{!}{
        \begin{tabular}{lcccc}
        \toprule

        \multirow{2}{*}{\textbf{Method}} &
        \multicolumn{3}{c}{\textbf{Pairwise Comparison}} &
        \multirow{2}{*}{\begin{tabular}[c]{@{}c@{}} \textbf{Publishable} \\ \textbf{Rate} \end{tabular} ($\uparrow$)} \\

        \cline{2-4}
        
        & 
        \multirow{1.25}{*}{\textbf{Win Rate} ($\uparrow$)} &
        \multirow{1.25}{*}{\textbf{Bradley--Terry} ($\uparrow$)} &
        \multirow{1.25}{*}{\textbf{Elo Rating} ($\uparrow$)} & \\
        
        \midrule
        
        Author-created                        & 0.157 & -0.503 & 1445 & 0.180 \\
        NanoBanana-Pro~\cite{nanobanana}          & 0.243 & \ 0.060 & 1489 & 0.377 \\
        PaperBanana~\cite{zhu2026paperbanana} & 0.357 & -0.683 & 1395 & 0.310 \\
        AutoFigure~\cite{zhu2026autofigure}   & 0.177 & -0.064 & 1442 & 0.223 \\
        GenGA (ours)                          & \ \textbf{0.767} & \textbf{1.189} & \textbf{1728} & \textbf{0.667} \\
        
        \bottomrule
        
        \end{tabular}
}
\end{table}

%% file: fig/tex/04_results.tex
\begin{figure}[!t]
    \centering
    \includegraphics[width=0.95\linewidth]{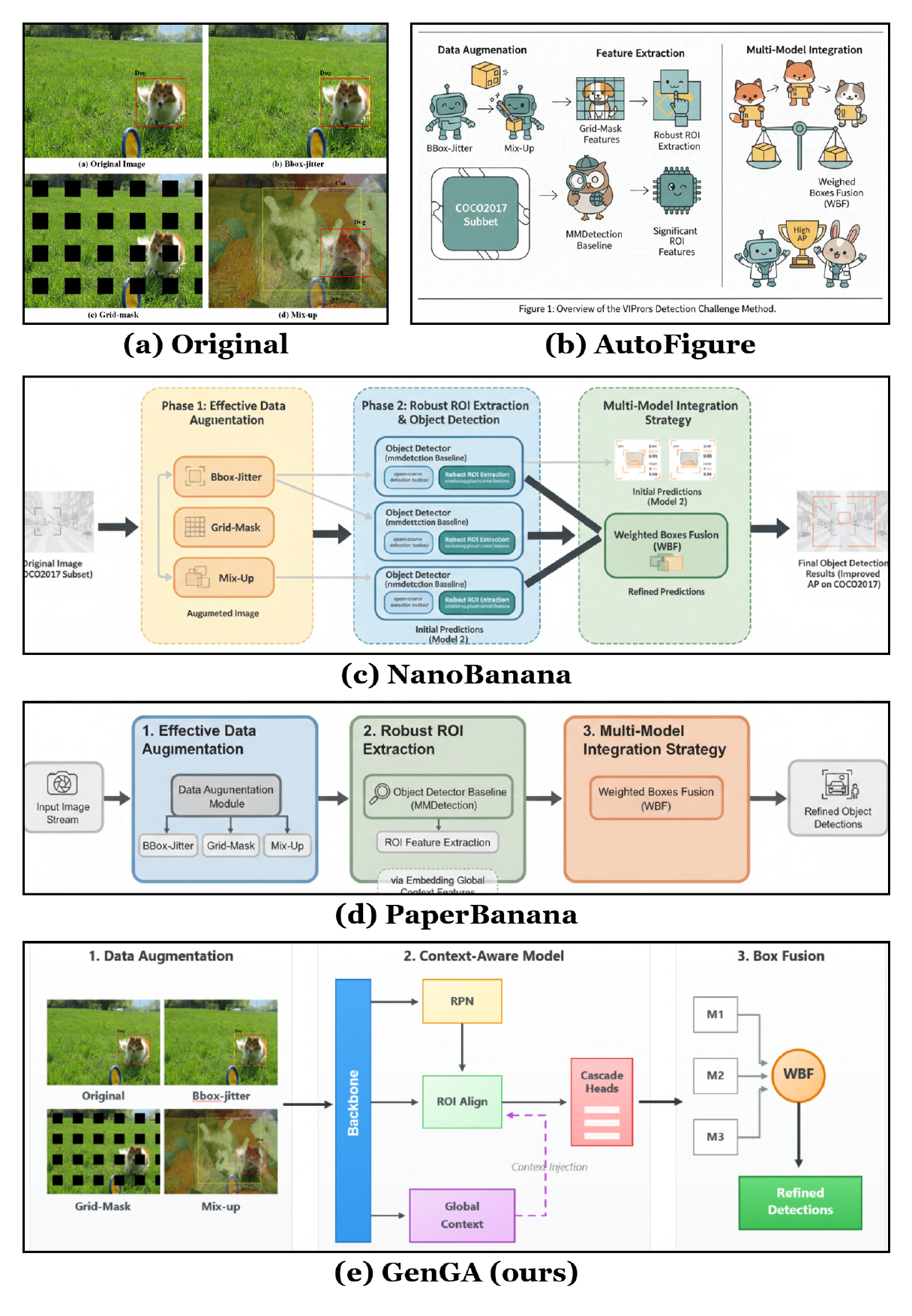}
    
    \caption{
        Qualitative comparison of generated GAs.
        Our method generates clean, well-structured vector representations with clearly separated regions, while avoiding common issues such as excessive icon usage, truncated or misaligned elements.
        (Source paper: \href{https://arxiv.org/abs/2104.09059}{10.48550/arXiv.2104.09059})
    }
    \label{fig:results}
\end{figure}

%% file: tbl/04_ablation.tex
\begin{table*}[!t]
\caption{
Ablation on reference input format and self-correction loop.
Combining raster and vector references with iterative refinement yields the best overall performance.
The best results for each metric are highlighted in \textbf{bold}.
}
\label{tbl:ablation}
\centering
\small
\resizebox{\textwidth}{!}{
\begin{tabular}{cccccccccccc}

    \toprule
    
    \multicolumn{2}{c}{\textbf{Reference Input}} & 
    \multirow{2}{*}{\begin{tabular}[c]{@{}c@{}} \textbf{Self-Correction} \\ \textbf{Loop} \end{tabular}} & 
    \multirow{2}{*}{\textbf{CLIP-S} ($\uparrow$) } &
    \multirow{2}{*}{\begin{tabular}[c]{@{}c@{}} \textbf{Overlap} \\ \textbf{Ratio}\end{tabular} ($\downarrow$) } &
    \multirow{2}{*}{\textbf{SIC} ($\uparrow$) } &
    \multirow{2}{*}{\begin{tabular}[c]{@{}c@{}} \textbf{Contribution QA } \\ \textbf{Accuracy}$^\dagger$ \end{tabular} ($\uparrow$)}  &
    \multicolumn{5}{c}{\textbf{VLM-as-a-Judge}$^\dagger$}\\
    
    \cline{1-2} \cline{8-12}
    
    \multirow{1.25}{*}{\textbf{Raster}} &
    \multirow{1.25}{*}{\textbf{Vector}} & & & & & &
    \multirow{1.25}{*}{\textbf{Faithfulness} ($\uparrow$)} &
    \multirow{1.25}{*}{\textbf{Conciseness} ($\uparrow$)} &
    \multirow{1.25}{*}{\textbf{Readability} ($\uparrow$)} &
    \multirow{1.25}{*}{\textbf{Aesthetics} ($\uparrow$)} &
    \multirow{1.25}{*}{\textbf{Overall} ($\uparrow$)}  \\
    
        \midrule
        
        \xmark & \xmark & \xmark &        0.230 & \textbf{0.116}&        0.948 &        0.377 / 0.422  &         0.432 / 0.203  &        0.387 / 0.532 &        0.552 / 0.230 &        0.417 /         0.165 &         0.380 / 0.177 \\
        \cmark & \xmark & \xmark &        0.230 &         0.127 &        0.954 &        0.387 / 0.434  &         0.397 / 0.123  &        0.396 / 0.533 &        0.560 / 0.257 &        0.475 /         0.245 &         0.359 / 0.230 \\
        \xmark & \cmark & \xmark &        0.253 &         0.166 &        0.955 &        0.407 / 0.427  &         0.391 / 0.132  &        0.403 / 0.578 &        0.549 / 0.242 &        0.452 /         0.183 &         0.363 / 0.205 \\
        \cmark & \cmark & \xmark &        0.254 &         0.141 &\textbf{0.959}&        0.392 / 0.429  &         0.400 / 0.150  &        0.391 / 0.563 &        0.569 / 0.257 &        0.441 /         0.228 &         0.381 / 0.252 \\
        \xmark & \xmark & \cmark &        0.249 &         0.170 &        0.926 &        0.357 / 0.428  & \textbf{0.439 / 0.208} &        0.382 / 0.567 &        0.648 / 0.340 &        0.522 /         0.333 &         0.399 / 0.322 \\
        \cmark & \xmark & \cmark &        0.249 &         0.190 &        0.931 &        0.392 / 0.450  &         0.402 / 0.172  &        0.382 / 0.580 &        0.638 / 0.328 &        0.666 / \textbf{0.442}&         0.385 / 0.307 \\
        \xmark & \cmark & \cmark &        0.252 &         0.178 &        0.936 &        0.372 / 0.416  &         0.403 / 0.175  &        0.443 / 0.645 &        0.669 / 0.378 &        0.612 /         0.370 &         0.403 / 0.337 \\
        \cmark & \cmark & \cmark &\textbf{0.269}&         0.163 &        0.943 &\textbf{0.422 / 0.467} &         0.427 / 0.198  &\textbf{0.473 / 0.690}&\textbf{0.683 / 0.392}&\textbf{0.690}/         0.438 & \textbf{0.473 / 0.362}\\
        
        \bottomrule
    
    \end{tabular}
}

\vspace{2pt}
\hfill {\footnotesize $\dagger$ Scores before/after the slash are reported using GPT-5.2 / Gemini-3.1-Pro-Preview, respectively, as the evaluator.}

\end{table*}

%% file: fig/tex/05_ablation.tex
\begin{figure}[!t]
    \centering
    \includegraphics[width=\linewidth]{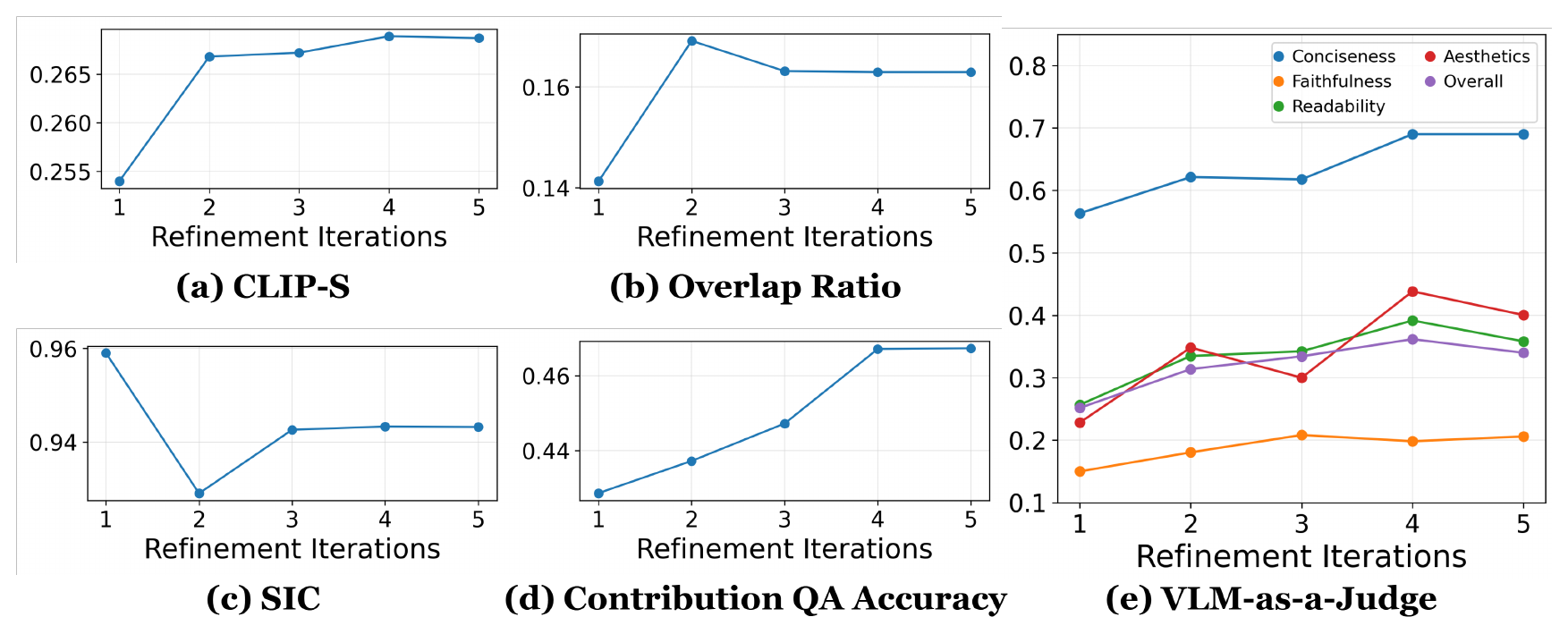}
    \caption{
        Effect of self-correction loop.
        Performance improves across multiple metrics, including semantic alignment, readability, and contribution understanding, as the number of refinement iterations increases, and stabilizes after around four iterations.
    }
    \label{fig:ablation}
\end{figure}

%% file: sec/06_conclusion.tex
\section{Conclusion}

We introduced Editable GA Generation, reframing GA generation as a structural, vector-based problem for human editing, along with SIC, a metric that captures editing simplicity and correlates with real editing cost.
We further proposed GenGA, a vector-first, data-grounded framework that generates GAs with superior editability and quality.
This work establishes GA generation a foundation for grounded in real research workflows and enhances scientific communication.

\paragraph{limitation.}
GenGA relies on high-capacity proprietary VLMs, raising concerns about cost, accessibility, and privacy due to external processing of potentially confidential data. Future enhancements could focus on directly optimizing editability and aesthetics, and enabling comparable performance with open, efficient local models.

%% file: sec/appendix/01_dataset_curation.tex
\section{Extraction and Curation of Asset Images}
\label{app:dataset_curation}

This section provides additional details on how visual assets were constructed from author-created GAs for dataset building and validation. Candidate asset regions were first extracted using LayerD. We then refined these candidates through rule-based preprocessing and manual curation.

In the preprocessing stage, regions that were extremely small relative to the original image were excluded from independent asset candidates, as they typically corresponded to text fragments, arrows, connectors, or other noisy components rather than reusable visual assets.

The final asset images were curated to satisfy four requirements: (i) semantic completeness, so that each asset preserved the main part of a depicted object as a coherent visual unit; (ii) proper separation from adjacent elements, so that visually distinct neighboring regions remained separated while each object was represented as a single natural unit; (iii) boundary quality, so that the crop did not include missing contours, excessive background, or contamination from adjacent elements; and (iv) annotation removal, so that text, arrows, label lines, symbols, and other auxiliary markings were excluded from the final crop. Candidate regions that did not satisfy these requirements were manually corrected by adjusting the crop, separating adjacent regions, or merging fragmented parts when necessary.

Figure~\ref{fig:dataset_curation} presents representative examples obtained through this procedure. The resulting assets are designed to serve as pseudo author-provided visual inputs, simulating the practical setting in which researchers supply images for downstream GA generation. We emphasize that this extraction pipeline was introduced only for dataset construction and validation; in the intended use of GenGA, visual assets are assumed to be directly provided by the authors.

%% file: sec/appendix/02_prompt.tex
\section{Prompt}
\label{app:prompt}

We provide the prompts used for the three VLM-based components in GenGA, namely, the Vectorizer, Generator, and Reviewer, in Prompt~\ref{prompt:vectorizer}, Prompt~\ref{prompt:generator}, and Prompt~\ref{prompt:reviewer}, respectively.
For the VLM-as-a-Judge evaluation, we follow PaperBanana and use exactly the same prompt and scoring rubric; therefore, we do not reproduce the judge prompt here.

%% file: sec/appendix/03_additional_results.tex
\section{Additional Results}
\label{app:additional_results}

Figure~\ref{fig:additional_results_a}, Figure~\ref{fig:additional_results_b}, and Figure~\ref{fig:additional_results_c} present additional qualitative examples illustrating how the self-correction loop progressively improves the generated GAs over refinement iterations. In early iterations, the outputs often contain ambiguous layouts, unclear labels, crossed arrows, collisions between visual elements, and occasional inconsistencies between the depicted structure and the paper content. As refinement proceeds, these issues are gradually resolved: layouts become more organized, labels become clearer, arrow routing and object placement are adjusted to reduce crossings and overlaps, and semantic inconsistencies in visual elements and connections are corrected.
Across examples, the final refined outputs are visually clearer, easier to interpret, and better aligned with the content of the source papers. Importantly, because GenGA produces figures in an editable vector format, these improvements are achieved while preserving element-level editability, allowing further manual refinement when needed.
In addition, when using Gemini-3.1-Pro-Preview, the average cost and processing time per paper are approximately \$0.80 and 8\,min\,7\,s, respectively.

%% file: sec/appendix/04_user_study_setup.tex
\section{User Study Setup and Participant Demographics}
\label{app:user_study_setup}

We provide additional details on the user study setup.
The study consisted of a figure-editing task conducted in draw.io and an online questionnaire administered via Google Forms.
Figure~\ref{fig:user_study_drawio} shows the editing interface used in the study, while Figure~\ref{fig:user_study_google_form} presents the questionnaire format and items.
Before the main study, participants completed a short practice session using draw.io to familiarize themselves with the editing interface and task format.
The practice examples were separate from those used in the main evaluation.
A total of 15 participants took part in the study: 4 master’s students, 3 researchers holding master’s degrees, 3 Ph.D. students, and 5 researchers holding Ph.D. degrees.
All participants had experience creating GAs or teaser figures and had published in peer-reviewed venues.

%% file: sec/appendix/_materials.tex
\onecolumn
\clearpage

\input{fig/tex/06_dataset_curation}
\clearpage

\begin{promptbox}{Prompt for the Vectorizer VLM, $f_{\text{vectorize}}(\cdot)$}{prompt:vectorizer}
\begin{PromptVerb}
# SYSTEM ROLE
- You are a professional scientific illustrator.

# TASK
- Convert the input image directly into a structured, semantic SVG.

# GUIDELINES

## STRICT CANVAS RULES
- The SVG canvas size and aspect ratio must EXACTLY match the input image.
- The root SVG element MUST be:
  <svg width="{W}" height="{H}" viewBox="0 0 {W} {H}">
- Do NOT add margins, padding, scaling, or transforms on the root element.
- Do NOT change width, height, or viewBox under any circumstances.

## STRUCTURAL GROUPING RULES
- Group related visual elements using <g> elements.
- Each major semantic block (e.g., input, method, output, example, visualization) MUST be wrapped in a <g>.
- Each such <g> MUST include:
  - class indicating its type (e.g., "module input", "module method", "module output")
  - data-role indicating its semantic role (e.g., data-role="input")

## ARROW ELEMENT RULES
- Represent connections and flows using explicit arrow elements.
- Arrows MUST be grouped in <g class="arrow"> elements.
- Each arrow group MUST include semantic attributes:
  - data-role="flow"
  - data-from="<SOURCE_ROLE>"
  - data-to="<TARGET_ROLE>"
- Use arrowheads defined via <defs> and <marker>.

## DRAWING RULES
- Use ONLY SVG primitives: <svg>, <g>, <rect>, <circle>, <ellipse>, <path>, <line>, <polygon>, <text>, <defs>, <marker>.
- Do NOT embed raster images (no <image> tags).
- Use simple geometric shapes and avoid unnecessary detail.

## LOGOS / ICONS / PHOTOS (PLACEHOLDERS) RULES
- Do NOT redraw logos, icons, photos, or complex pictorial elements.
- Replace them with rectangular placeholders.
- Placeholders MUST:
  - Match the original position and size.
  - Use a neutral visual style (light gray fill, dashed stroke).
  - Be explicitly tagged.

## PLACEHOLDER TAGGING RULES:
- Every placeholder rectangle MUST include:
  - class="placeholder"
  - data-placeholder="true"
  - data-placeholder-type (e.g., "icon", "logo", "photo", "output_example", "visualization")
- If applicable, also include:
  - data-semantic-role describing its meaning within the figure.

## TEXT ELEMENT RULES
- Preserve all visible text verbatim.
- Place text at the correct relative positions.
- Prioritize text correctness over decoration.
- All text content MUST be valid XML.
- Escape &, <, > as &amp;, &lt;, &gt;.

## LAYOUT RULES
- Preserve spatial layout, alignment, and relative positioning.
- Maintain the logical structure of boxes, arrows, and groups.

# FINAL CHECK
- Ensure width, height, and viewBox are correct.
- Ensure every major element is inside a semantic <g>.
- Ensure all placeholders and arrows are correctly tagged.
- Compare the SVG against the original input image.
- Ensure that the size, shape, and relative proportions of all elements
  (boxes, arrows, placeholders, text blocks) visually match the original image.
- Ensure that no element is significantly enlarged, shrunk, or simplified
  compared to its appearance in the original image.
- Ensure that arrow lengths, directions, and connection points are visually consistent.
- If any visual mismatch is detected, adjust the SVG to better match the original image
  while preserving the defined semantic structure and tags.

# OUTPUT RULES
- Output SVG code ONLY.
- Do NOT include explanations, comments, or markdown.
\end{PromptVerb}
\end{promptbox}

\begin{promptbox}{Prompt for the Generator VLM, $f_{\text{generate}}(\cdot)$}{prompt:generator}
\begin{PromptVerb}
# SYSTEM ROLE
- You are GenGA, an automated system for generating graphical abstracts for scientific papers.
- A graphical abstract is a single clean visual summary of the main idea.
- A graphical abstract is typically used as Fig.1 or teaser figure.

---------------------------------

# TASK
- Generate a scientific graphical abstract as an SVG.
- The SVG must faithfully represent the paper content below.
- Aesthetic styling (color, shadows, gradients, decorations) is NOT required.

## PAPER CONTENT
```
{paper_content}
```

## OUTPUT FORMAT
- Output MUST be a ONLY single valid SVG.
- Do NOT include any text, explanation, or markdown outside the SVG.
```
<svg>
   ...
</svg>
```

# GUIDELINES

## GENERAL RULES
- Generate a concise and conceptually clear graphical abstract.
- Focus on the main idea, not detailed methodology or full experiments.
- Do NOT create a slide-style layout.
- Do NOT include the paper title inside the figure.
- Do NOT include keyword lists.

## LAYOUT RULES
- No overlap between exclusive siblings.
- Maintain consistent spacing between structural blocks.
- Align elements to an invisible grid.
- Avoid excessive empty space.

## HIERARCHY RULES
- All visual elements MUST be organized using nested <g> groups.
- Every element must belong to a clear parent group.
- No element should exist directly under <svg> without being inside a <g>.
- The hierarchy defines structural containment and geometric constraints.
- A child element MAY overlap with its parent.
- A child element MUST remain inside the spatial bounds of its parent.
- Sibling elements at the same hierarchy level must follow geometry policy rules.

## GEOMETRY POLICY LABELS
Each <g> element must include a geometry policy attribute:
1. geom-policy="exclusive"
- This element must NOT overlap with other 'exclusive' elements at the same hierarchy level.
- Used for structural layout blocks.
2. geom-policy="overlay"
- This element MAY overlap other elements at the same hierarchy level.
- It must still remain inside its parent bounds.
- Used for connectors, arrows, or minor annotations.

## TEXT ELEMENT RULES
- Use short labels only.
- Avoid full sentences.
- Avoid paragraph text.
- Text must be directly attached to related elements.

## ARROW ELEMENT RULES
- Arrows represent conceptual flow only.
- Prefer straight horizontal or vertical arrows.
- Avoid unnecessary bends or zigzag paths.
- Avoid diagonal arrows unless absolutely necessary.
- Arrowheads must not be excessively large.
- An arrow must NOT intersect unrelated exclusive structural blocks.
- If a straight arrow would intersect another element,
  adjust the layout instead of bending excessively.

## STRUCTURAL ELEMENT RULES
- You may generate geometric primitives to define structure (e.g., <rect>, <circle>, <line>, <path> (arrows only), <text>).
- These elements are allowed only for:
  - Region separation
  - Flow indication
  - Conceptual grouping
- Do not use geometric primitives to draw complex icons or images.

## NON-STRUCTURAL ELEMENT RULES
- The following MUST NOT be drawn using shapes or paths:
  - Icons or pictograms (e.g., lightbulb, brain, computer, robot, human, car, molecule, cell)
  - Example input/output images
  - Maps or geographic regions
  - Graphs, plots, scatter charts
- These elements are considered raster-type semantic objects.
- They must not be approximated using vector geometry.
- Instead, use placeholder groups explained bellow.

## PLACEHOLDER GROUP ELEMENT RULES
- Placeholders define layout slots for non-geometric content.
- A placeholder MUST follow this structure:
```
<g class="placeholder" geom-policy="exclusive">
    <rect class="placeholder-rect" fill="#eef1f6" stroke="#9aa3ad" />
    <text class="placeholder-label">{description}</text>
</g>
```
- The actual image will be inserted downstream.
- Do NOT attempt to visually approximate the object.
- Use placeholders for all provided assets bellow.

## ASSET RULES
- Asset images are visual materials (e.g., icons, example illustrations, style primitives)
- Assets use the same placeholder structure as regular placeholders.
- The only difference is the required attribute: asset-id="ASSET_ID".
- Placeholder width and height MUST exactly match the provided values.
- All assets MUST be used.
- However, if there are assets with truly unclear intent or that are extremely difficult to place in the layout, as a last resort, it is acceptable to not use them.
```
<g class="placeholder" asset-id="{ASSET_ID}" geom-policy="exclusive">
    <rect class="placeholder-rect" fill="#eef1f6" stroke="#9aa3ad" />
    <text class="placeholder-label">"{description}"</text>
</g>
```

# REFERENCE VECTOR GRAPHICAL ABSTRACT INFO:
- The SVG provided below is the reference image.
- Reference SVG is for layout and structural inspiration only.
- Do NOT copy paths, text, or coordinates verbatim.
- Do NOT trace or replicate path-level details.
```
{vector_reference_svg}
```

# REFERENCE RASTER GRAPHICAL ABSTRACT INFO:
- The first image provided above is reference image.
- Reference image is for layout/structure inspiration only.
- Do NOT copy text, icons, colors, or visual style.
- Do NOT trace or replicate pixel-level details.

# PROVIDED ASSETS INFO:
- Asset images are provided above in order.
- The j-th asset image corresponds to the j-th description below.
- NOTE: a raster reference image is also provided above; the FIRST image is a reference and NOT an asset.
- asset_id: {asset_id_1}:
  - description: {asset_description_1}
  - width: {asset_width_1}px
  - height: {asset_height_1}px
- asset_id: {asset_id_2}:
  - description: {asset_description_2}
  - width: {asset_width_2}px
  - height: {asset_height_2}px
- asset_id: {asset_id_3}:
  - description: {asset_description_3}
  - width: {asset_width_3}px
  - height: {asset_height_3}px
\end{PromptVerb}
\end{promptbox}

\begin{promptbox}{Prompt for the Reviewer VLM, $f_{\text{review}}(\cdot)$}{prompt:reviewer}
\begin{PromptVerb}
# SYSTEM ROLE
- You are GenGA, an automated system for generating graphical abstracts for scientific papers.
- A graphical abstract is a single clean visual summary of the main idea.
- A graphical abstract is typically used as Fig.1 or teaser figure.

# TASK
- Review the overall quality of the generated graphical abstract.
- Focus on high-level structure and conceptual clarity.
- The raster graphic above is the PRIMARY object under review.
- The SVG below is provided for structural reference.
- Provide constructive feedback on the following aspects:
   - Faithfulness to the core idea of the paper
   - Conceptual clarity
   - Logical flow of information
   - Balance and visual composition
   - Overall readability and effectiveness as a graphical abstract

## PAPER CONTENT
```
{paper_content}
```

## PREVIOUS GENERATED SVG
```
{previous_generated_svg}
```
\end{PromptVerb}
\end{promptbox}
\clearpage

\input{fig/tex/07_additional_results_a}
\input{fig/tex/08_additional_results_b}
\input{fig/tex/09_additional_results_c}

\input{fig/tex/10_user_study_drawio}

\input{fig/tex/11_user_study_google_form}

%% file: fig/tex/06_dataset_curation.tex
\begin{figure}[!t]
    \centering
    \begin{minipage}{\linewidth}
        \centering
        \includegraphics[width=\textwidth]{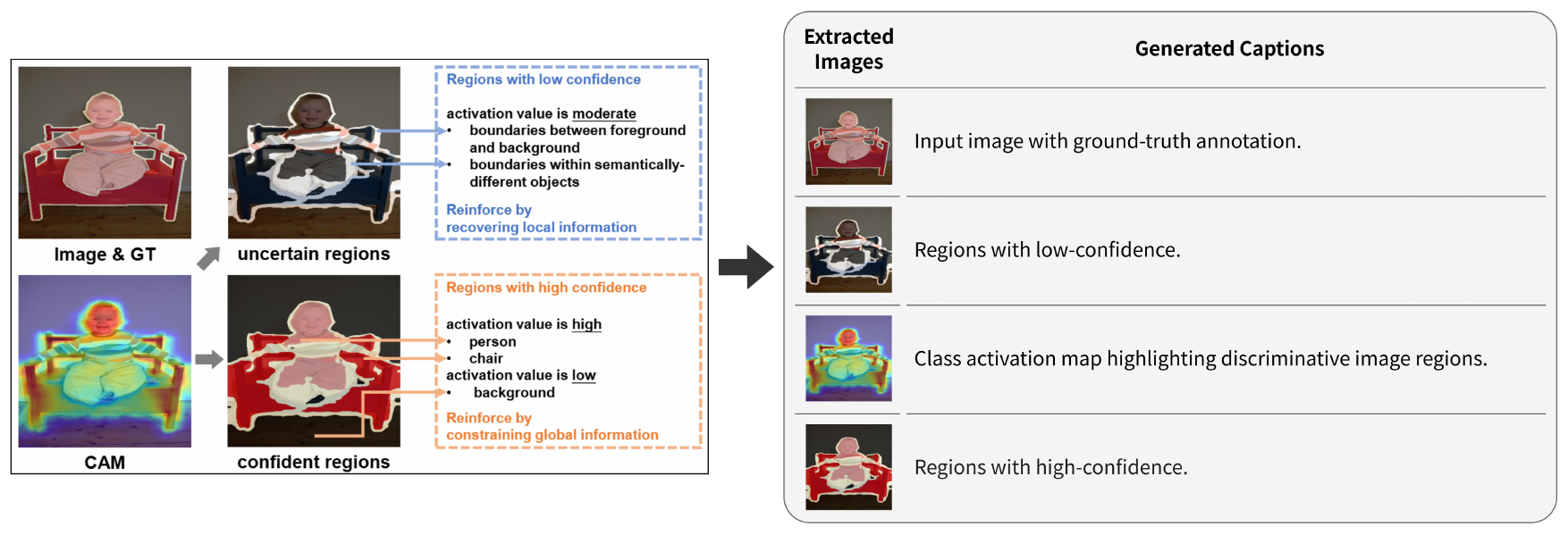}
        \subcaption{Source paper: \href{https://arxiv.org/abs/2312.08916}{10.48550/arXiv.2312.08916}}
        \label{fig:dataset_curation:a}
    \end{minipage}   
    \begin{minipage}{\linewidth}
        \centering
        \includegraphics[width=\textwidth]{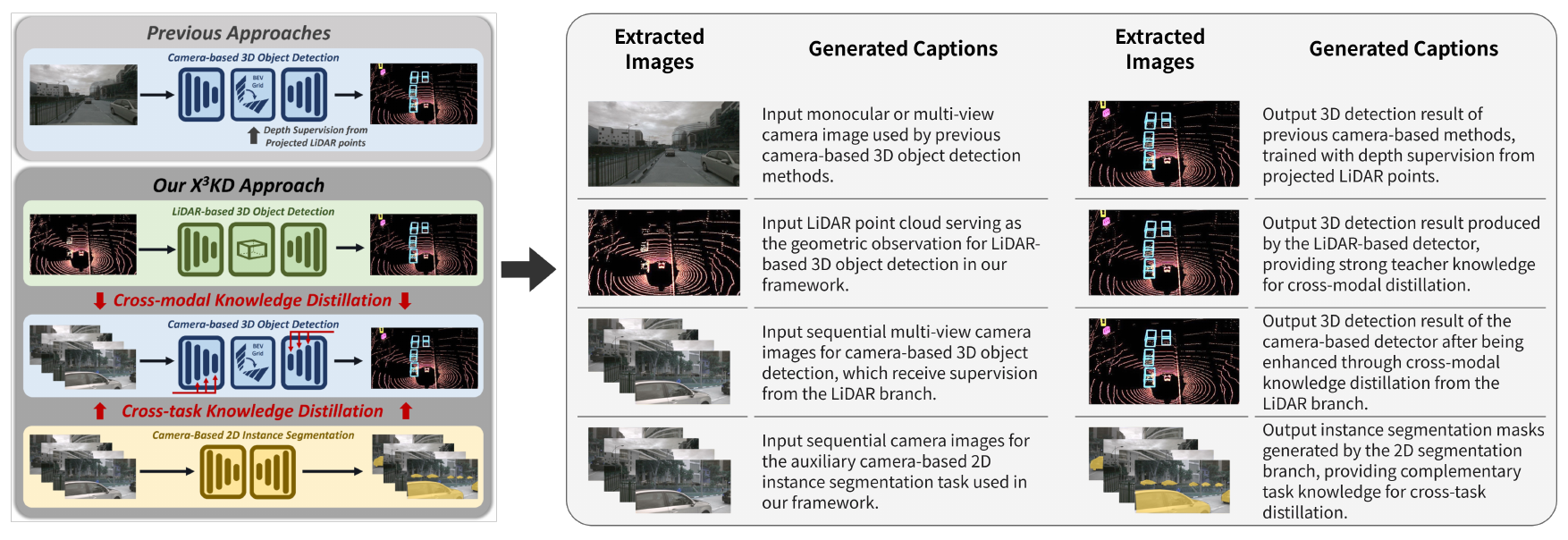}
        \subcaption{Source paper: \href{https://arxiv.org/abs/2303.02203}{10.48550/arXiv.2303.02203}}
        \label{fig:dataset_curation:b}
    \end{minipage}
    \caption{
        Examples of extracted asset images.
        For each source paper, candidate visual regions are extracted from author-created GAs, refined to preserve semantically coherent objects while removing annotations and noisy boundaries, and paired with automatically generated captions.
        These curated asset images are used as pseudo author-provided raw visual inputs, simulating the real-world use case in which researchers supply images to be incorporated into generated GA.
    }
    \label{fig:dataset_curation}
\end{figure}

%% file: fig/tex/07_additional_results_a.tex
\begin{figure}
    \centering
    \includegraphics[width=\linewidth]{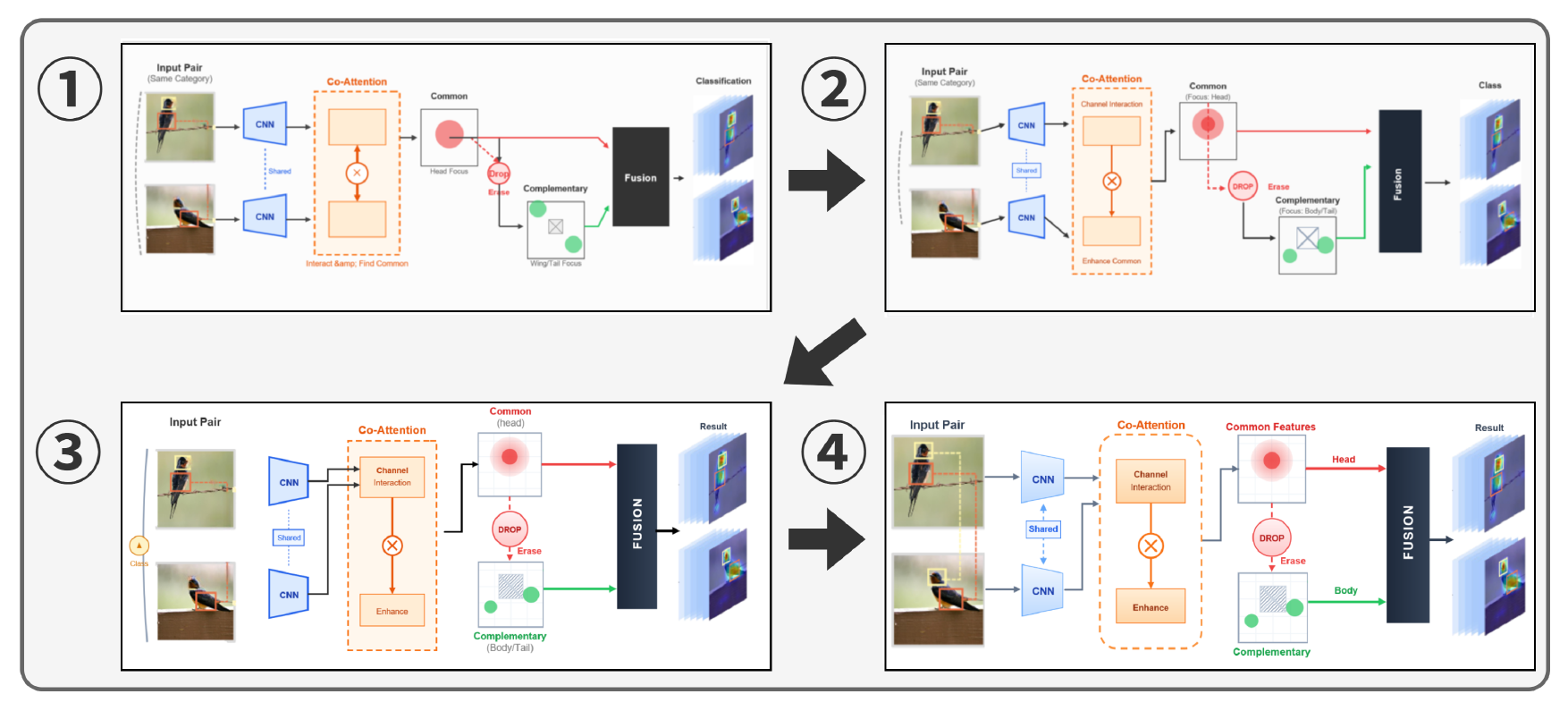}
    \caption{
        Additional qualitative example showing how the self-correction loop progressively improves the generated GA over four refinement iterations.
        Early iterations exhibit ambiguous layout structure, weak alignment between labels and modules, and less organized visual flow, while later iterations produce clearer grouping, better arrow routing, and improved semantic consistency with the source paper.
        (Source paper: \href{https://arxiv.org/abs/2101.08527}{10.48550/arXiv.2101.08527})
    }
    \label{fig:additional_results_a}
\end{figure}

%% file: fig/tex/08_additional_results_b.tex
\begin{figure}
    \centering
    \includegraphics[width=\linewidth]{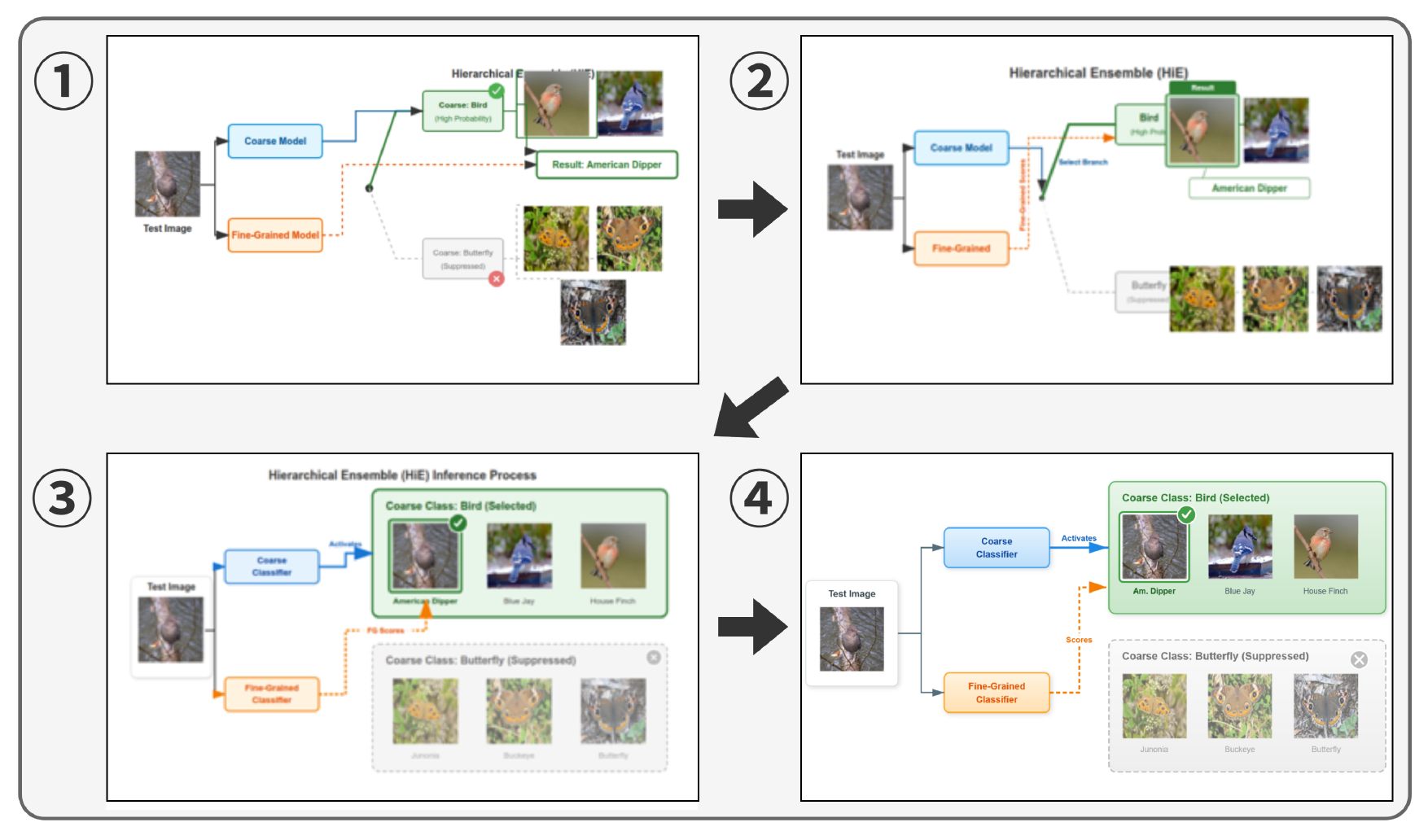}
    \caption{
        Additional qualitative example showing how the self-correction loop progressively improves the generated GA over four refinement iterations.
        As the number of iterations increases, the GA becomes more structured and readable, with improved placement of example images, clearer hierarchical organization, and reduced visual clutter, resulting in a representation that better matches the paper’s core idea. 
        (Source paper: \href{https://arxiv.org/abs/2302.00368}{10.48550/arXiv.2302.00368})
    }
    \label{fig:additional_results_b}
\end{figure}

%% file: fig/tex/09_additional_results_c.tex
\begin{figure}
    \centering
    \includegraphics[width=\linewidth]{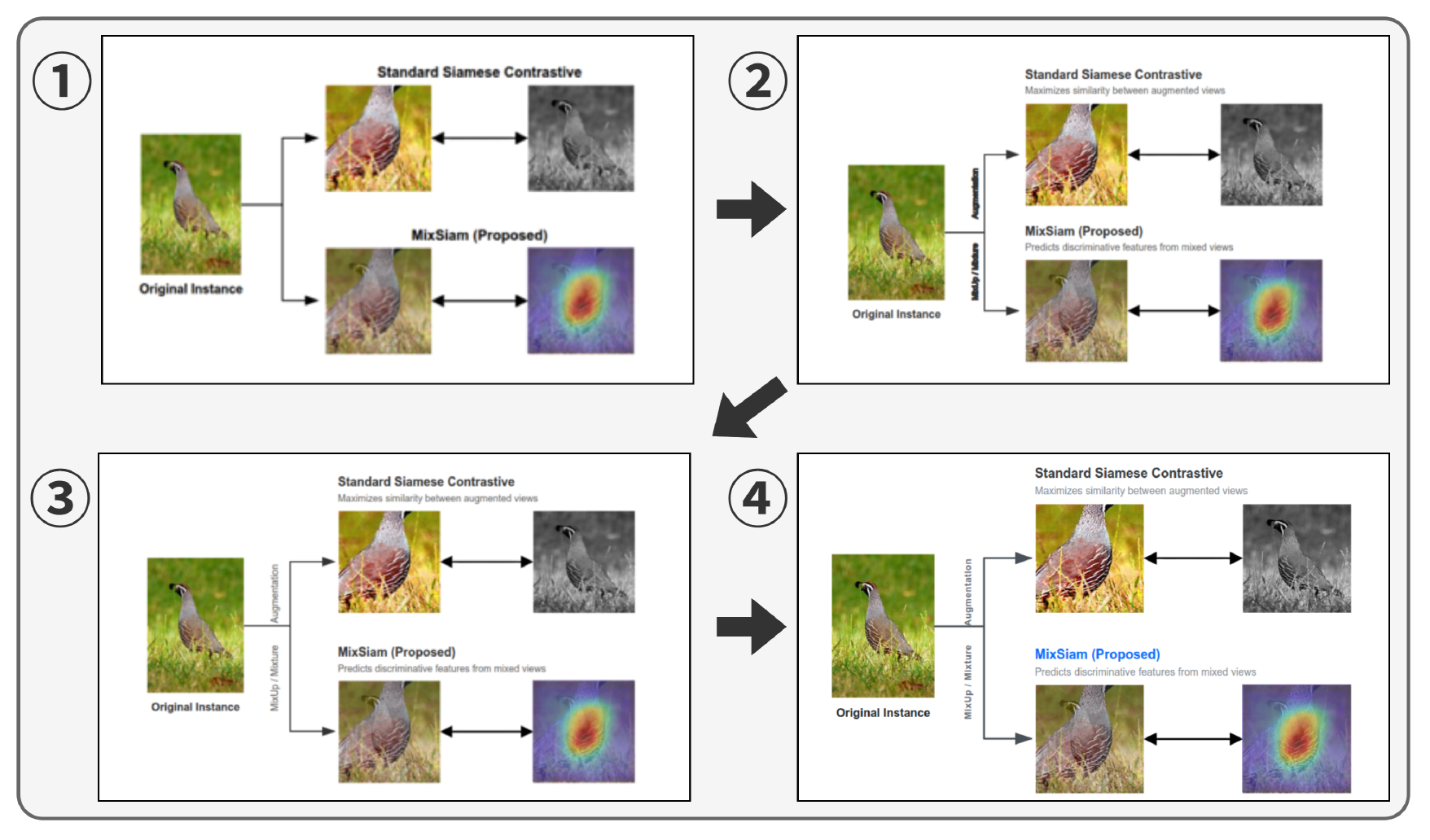}
    \caption{
        Additional qualitative example showing how the self-correction loop progressively improves the generated GA over four refinement iterations.
        The self-correction loop improves text clarity, visual balance, and semantic correspondence between the contrasted views and explanatory modules, leading to a more polished and interpretable final result while preserving element-level editability.
        (Source paper: \href{https://arxiv.org/abs/2111.02679}{10.48550/arXiv.2111.02679})
    }
    \label{fig:additional_results_c}
\end{figure}

%% file: fig/tex/10_user_study_drawio.tex
\begin{figure}[!t]
    \centering
    \includegraphics[width=\linewidth]{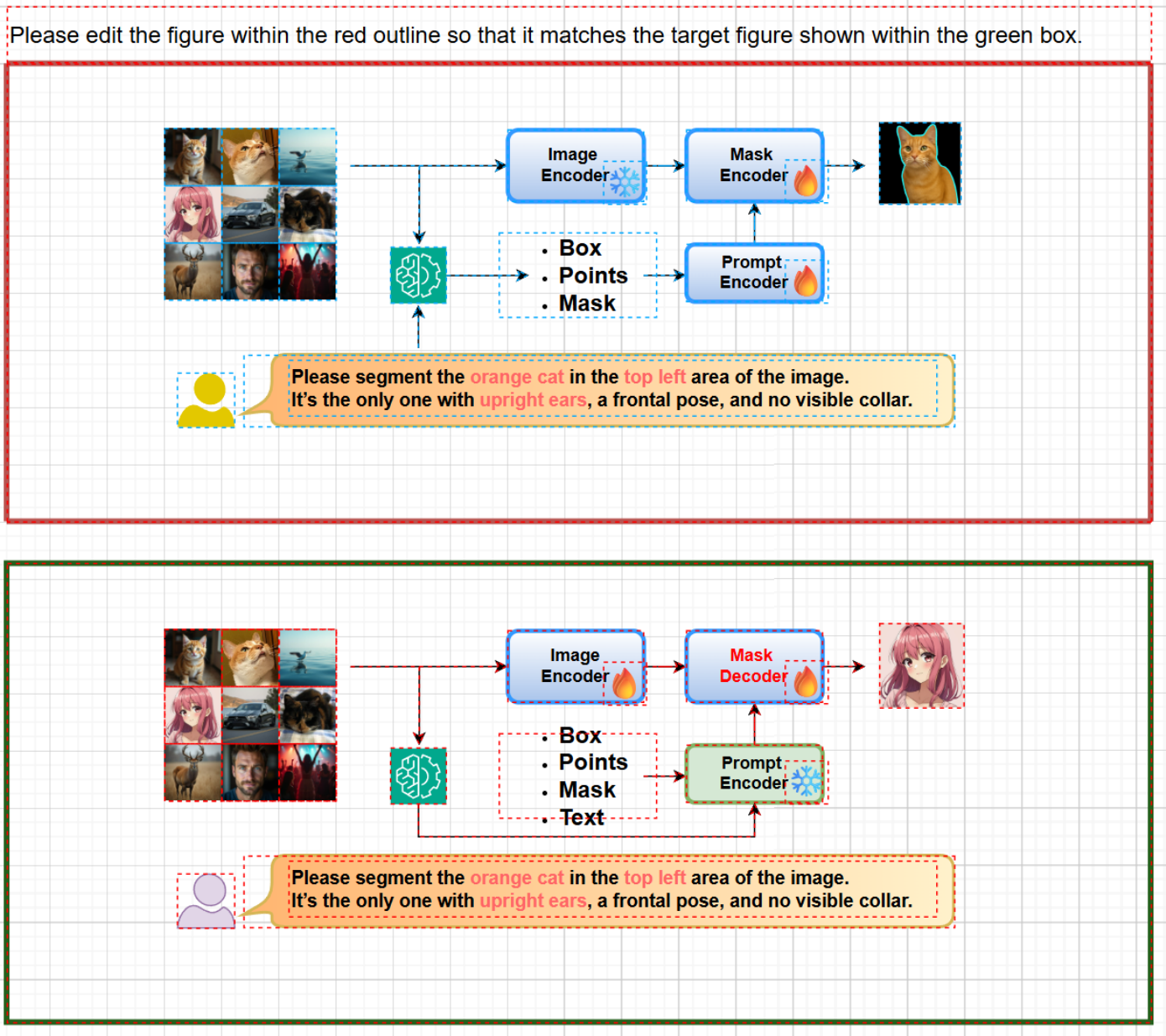}
    
    \caption{
        Editing interface used in the figure-editing task of the user study. Participants were asked to modify the figure inside the red outline so that it matched the target figure shown in the green box, using draw.io within a fixed time budget.
    }
    \label{fig:user_study_drawio}
\end{figure}

%% file: fig/tex/11_user_study_google_form.tex
\begin{figure}[!t]
    \centering

    \begin{minipage}{0.49\linewidth}
        \centering
        \includegraphics[width=\textwidth]{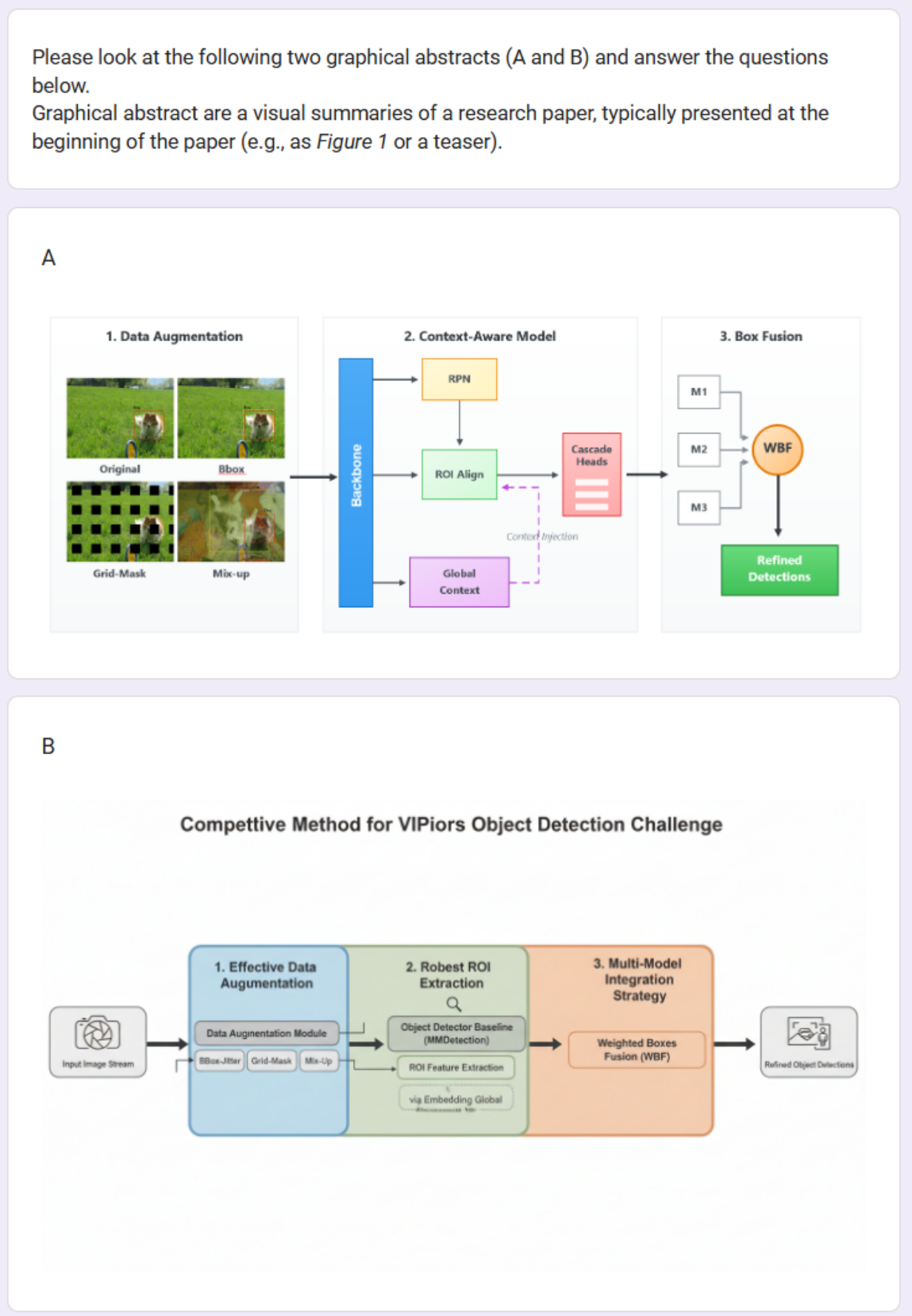}
        \label{fig:user_study_google_form:a}
    \end{minipage}   
    \begin{minipage}{0.49\linewidth}
        \centering
        \includegraphics[width=\textwidth]{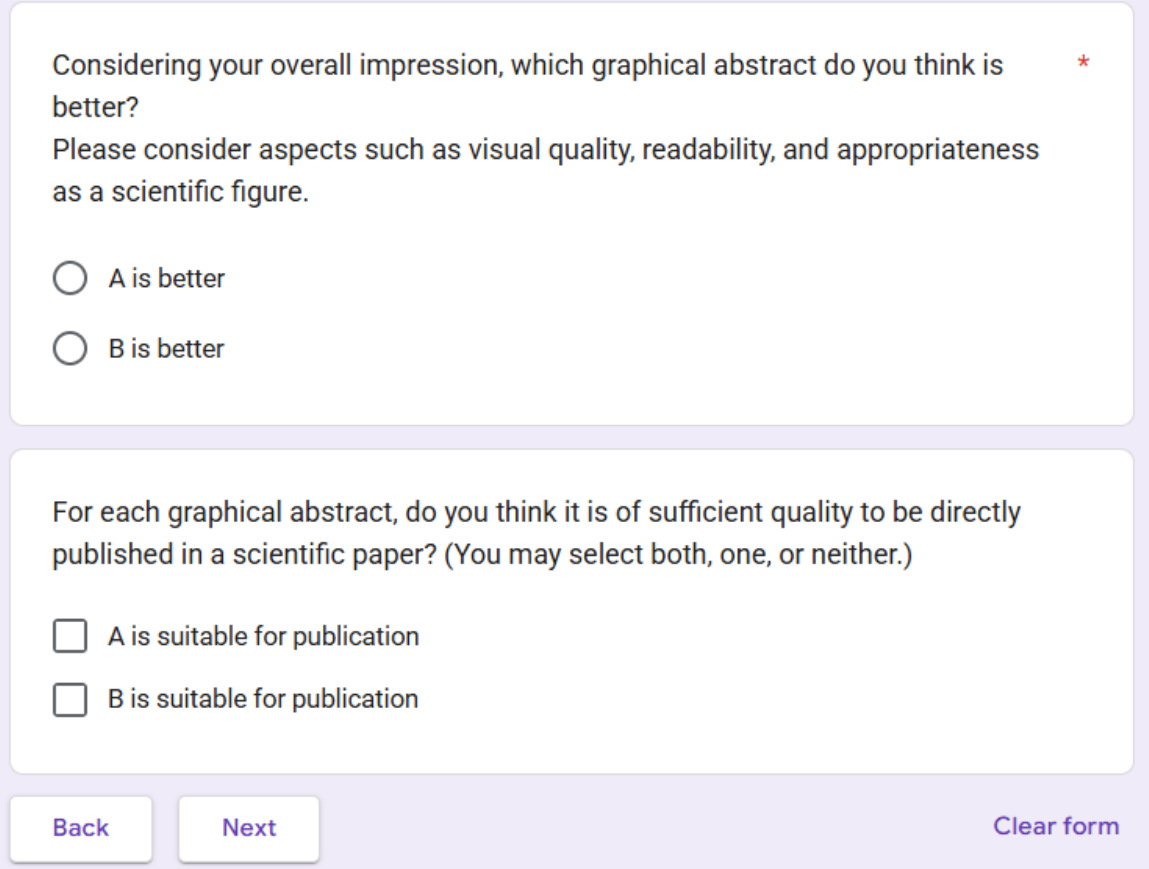}
        \label{fig:user_study_google_form:b}
    \end{minipage}

    \caption{
        Online questionnaire used in the user study. Participants compared pairs of GAs, selected the overall preferred one, and independently judged whether each GA was of sufficient quality to be directly published in a scientific paper.
    }
    \label{fig:user_study_google_form}
\end{figure}